\input harvmac.tex
\overfullrule=0mm

\input epsf.tex
\newcount\figno
\figno=0
\def\fig#1#2#3{
\par\begingroup\parindent=0pt\leftskip=1cm\rightskip=1cm\parindent=0pt
\baselineskip=11pt
\global\advance\figno by 1
\midinsert
\epsfxsize=#3
\centerline{\epsfbox{#2}}
\vskip 12pt
{\bf Fig. \the\figno:} #1\par
\endinsert\endgroup\par
}
\def\figlabel#1{\xdef#1{\the\figno}}
\def\encadremath#1{\vbox{\hrule\hbox{\vrule\kern8pt\vbox{\kern8pt
\hbox{$\displaystyle #1$}\kern8pt}
\kern8pt\vrule}\hrule}}


%
\def\frac#1#2{\scriptstyle{#1 \over #2}}

\def\vev#1{\langle #1 \rangle}
\def\ket#1{ | #1 \rangle}
\def\bra#1{ \langle #1 |}

%
%
\def\CA{{\cal A}}		\def\CC{{\cal C}}
\def\CD{{\cal D}}		
\def\CG{{\cal G}}	\def\CH{{\cal H}}	
		
	\def\CN{{\cal N}}	
\def\CP{{\cal P}}		\def\CR{{\cal R}}
		\def\CU{{\cal U}}
\def\CV{{\cal V}}		

\def\({ \left( }\def\[{ \left[ }
\def\){ \right) }\def\]{ \right] }
%


\def\IR{\relax{\rm I\kern-.18em R}}
\font\cmss=cmss10 \font\cmsss=cmss10 at 7pt
\def\IZ{\relax\ifmmode\mathchoice
{\hbox{\cmss Z\kern-.4em Z}}{\hbox{\cmss Z\kern-.4em Z}}
{\lower.9pt\hbox{\cmsss Z\kern-.4em Z}}
{\lower1.2pt\hbox{\cmsss Z\kern-.4em Z}}\else{\cmss Z\kern-.4em Z}\fi}
\def\inbar{\,\vrule height1.5ex width.4pt depth0pt}
\def\IB{\relax{\rm I\kern-.18em B}}
\def\IC{\relax\hbox{$\inbar\kern-.3em{\rm C}$}}
\def\ID{\relax{\rm I\kern-.18em D}}
\def\IE{\relax{\rm I\kern-.18em E}}
\def\IF{\relax{\rm I\kern-.18em F}}
\def\IG{\relax\hbox{$\inbar\kern-.3em{\rm G}$}}
\def\IH{\relax{\rm I\kern-.18em H}}
\def\II{\relax{\rm I\kern-.18em I}}
\def\IK{\relax{\rm I\kern-.18em K}}
\def\IL{\relax{\rm I\kern-.18em L}}
\def\IM{\relax{\rm I\kern-.18em M}}
\def\IN{\relax{\rm I\kern-.18em N}}
\def\IO{\relax\hbox{$\inbar\kern-.3em{\rm O}$}}
\def\IP{\relax{\rm I\kern-.18em P}}
\def\IQ{\relax\hbox{$\inbar\kern-.3em{\rm Q}$}}
\def\IGa{\relax\hbox{${\rm I}\kern-.18em\Gamma$}}
\def\IPi{\relax\hbox{${\rm I}\kern-.18em\Pi$}}
\def\ITh{\relax\hbox{$\inbar\kern-.3em\Theta$}}
\def\IOm{\relax\hbox{$\inbar\kern-3.00pt\Omega$}}



\def\oh{{1\over 2}}\def\un{{\bf 1}}

\def\Ga{\alpha}\def\Gb{\beta}\def\Gc{\gamma}\def\GC{\Gamma}
\def\Gd{\delta}\def\GD{\Delta}\def\Ge{\epsilon}

\def\Gl{\lambda}\def\GL{\Lambda}
\def\Gm{\mu}\def\Gn{\nu}\def\ksi{\xi}
\def\Gr{\rho}
\def\Gs{\sigma}\def\Gt{\tau}


\def\mod{{\rm mod\,}}
\def\diag{{\rm diag \,}}
 
\def\bra{\langle}\def\ket{\rangle}
\def\nind{\par\noindent}

 \def\Che{Chebishev\ }
 
\def\ie{{\it i.e.\ }}

\def\Exp{{\rm Exp}\,}  
\def\suh{\widehat{su}}
\def\hepth#1{{\tt hep-th #1}}

%
\def\msy{y }
\message{Do you have the AMS fonts (y/n) ?}\read-1 to \msan
\ifx\msan\msy
\input amssym.def
\input amssym.tex
\def\IZ{\Bbb Z}\def\IR{\Bbb R}\def\IC{\Bbb C}\def\IN{\Bbb N}

\else 

\fi
\Title{\vbox{\hbox{SPhT 95/089}
\hbox{{\tt hep-th/9507057   }}}}
{{\vbox {
\medskip
\centerline{Graphs and Reflection Groups} 
}}}

\bigskip

\centerline{J.-B. Zuber}\bigskip

\centerline{ \it CEA, Service de Physique Th\'eorique de Saclay
,}
\centerline{ \it F-91191 Gif sur Yvette Cedex, France}

\vskip .2in
\noindent 
It is shown that graphs that generalize the ADE Dynkin diagrams
and have appeared in various contexts of two-dimensional field theory
may be regarded in a natural way as encoding the geometry of a 
root system. After recalling what are the conditions 
satisfied by these graphs, we define a bilinear form on a root system  
in terms of the adjacency matrices of these graphs and 
undertake the study of the group generated by the
reflections in the hyperplanes orthogonal to these roots. 
Some ``non integrally laced " graphs are shown to be associated with 
subgroups of these reflection groups. The empirical 
relevance of these graphs
in the classification of conformal field theories or in the 
construction of integrable lattice models is recalled, and the
connections with recent developments 
in the context of ${\cal N}=2$ supersymmetric theories and 
topological field theories are discussed. 


\Date{
7/95\qquad\qquad   To be submitted to Communications in Mathematical Physics}
%


\lref\ICMP{J.-B. Zuber, {\it Conformal, Integrable and Topological Theories,
Graphs and Coxeter Groups}, SPhT 94/156, {\tt hep-th/9412202}, 
to appear in the proceedings of the International Conference of Mathematical
Physics, Paris July 1994, D. Iagolnitzer edr, International Press, Boston.}
\lref\Hum{J.E. Humphreys, {\it Reflection Groups and Coxeter Groups}, 
Cambridge Univ. Pr. 1990 \semi 
N. Bourbaki, {\it Groupes et Alg\`ebres de Lie}, chap. 4--6, Masson 1981.  }
\lref\Cox{H.S.M. Coxeter, {\it Ann. Math.} {\bf 35} (1934) 588. 
}
\lref\GHJ{F.M. Goodman, P. de la Harpe and V.F.R. Jones, 
{\it Coxeter Graphs and Towers of Algebras}, Springer Verlag 1989. }
\lref\Verl{E. Verlinde, {\it Nucl. Phys.} {\bf 300} (1988) [FS22] 360.} 
\lref\KMoPa{V. Kac, {\it Funct. Anal. and Appl.} {\bf 3} (1969) 252
\semi R.V. Moody and J. Patera, {\it SIAM J. Alg. Disc. Meth.} {\bf 5}
(1984) 359 
and further references therein. }
\lref\Gepn{D. Gepner, Comm. Math. Phys. {\bf 141 } (1991) 381. } 
\lref\DFZun{P. Di Francesco and J.-B. Zuber, 
{\it Nucl. Phys.} {\bf B338} (1990) 602
.} 
\lref\Coxe{H.S.M. Coxeter, {\it Duke Math. J.} {\bf 18} (1951) 765.} 
\lref\KS{Y. Kazama and H. Suzuki, {\it Phys. Lett.} {\bf B216} (1989) 112; 
{\it Nucl. Phys.} {\bf B321} (1989) 232. }
\lref\DFLZ{
D. Nemeschansky, and N.P. Warner, Nucl. Phys. {\bf B380} (1992) 241
\semi 
P. Di Francesco, F. Lesage and J.-B. Zuber,
Nucl. Phys. {\bf B408} (1993) 600.}
\lref\CoMo{ H.S.M. Coxeter and W.O.J. Moser, 
{\it Generators and Relations for Discrete Groups}, Springer 1957. }
\lref\Fu{J. Fuchs, {\it Fusion rules in conformal field theory}, 
\hepth{9306162} and further references therein.}
\lref\KT{
A. Kuniba and T. Nakanishi, {\it Level-Rank Duality in Fusion RSOS Models}
proceedings of the International Colloquium on Modern Quantum Field Theory,
January 1990, Bombay, India.}
\lref\AGZV{V.I. Arnold,  and S.M. Gusein-Zade, A.N. Varchenko,
{\it Singularities of differentiable maps}, Birk\"auser, Basel 1985.}
\lref\Saito{K. Saito, {\it Adv. Stud. Pure Math.} {\bf 8} (1986) 479-526.}
\lref\AC{N. A'Campo, {\it Math. Ann.} {\bf 213} (1975) 1.}
\lref\Nick{N. Warner, private communication.}
\lref\Zub{J.-B. Zuber, {\it Mod. Phys. Lett. A} {\bf 8} (1994) 749. }
\lref\VP{V. Pasquier, J.Phys. {\bf A20} (1987)  5707 } 
%
\lref\Ocun{A. Ocneanu,  
in {\it Operator Algebras and Applications}, vol.2, 119-172, London Math. Soc.
Lecture Notes Series, Vol. 136, Cambridge Univ. Press, London 1988.}
\lref\SMP{O.P. Shcherbak, {\it Russ. Math. Surveys} 
{\bf 43:3}  (1988) 149 
\semi R.V. Moody and J. Patera, {J.Phys.A} {\bf 26} (1993) 2829.} 
\lref\ChRa{P. Christe and F. Ravanini, {\it Int. J. Mod. Phys.} 
{\bf A4} (1989) 897. }
\lref\VPun{V. Pasquier, {\it Nucl. Phys.} {\bf B285} [FS19] (1987) 162.} 
\lref\DFZ{P. Di Francesco and J.-B. Zuber, 
in {\it Recent Developments in Conformal Field Theories}, Trieste
Conference, 1989, S. Randjbar-Daemi, E. Sezgin and J.-B. Zuber eds., 
World Scientific 1990
\semi P. Di Francesco, {\it Int.J.Mod.Phys.} {\bf A7} 407-500 (1992).}
\lref\CIZ{A. Cappelli, C. Itzykson, and J.-B. Zuber,
{\it Nucl. Phys.} {\bf B280} (1987) [FS]
445; {\it Comm. Math. Phys.} {\bf 113} (1987) 1
\semi A. Kato, {\it Mod. Phys. Lett.} {\bf A2} (1987) 585.}
\lref\Itz{C. Itzykson, {\it Nucl. Phys. Proc. Suppl.} {\bf 58 } (1988) 150.}
\lref\Gan{ T. Gannon, 
{\it Comm. Math. Phys.} {\bf 161} (1994) 233; 
{\it The classification of SU(3) modular invariants revisited},  
{\tt hep-th 9404185}. }
\lref\PZ{V. Petkova and J.-B. Zuber, 
{\it Nucl. Phys. B}. {\bf B438} (1995) 347. }
\lref\PZnew{V. Petkova and J.-B. Zuber, 
{\it From CFT's to graphs} to appear. } 
\lref\BI{E. Bannai and T. Ito, 
{\it Algebraic Combinatorics I: Association schemes}, Benjamin-Cummins 
1984.}
\lref\Ocn{A. Ocneanu, communication at the Workshop
{\it Low Dimensional Topology, Statistical Mechanics and Quantum Field Theory},
 Fields Institute, Waterloo, Ontario, April 26--30, 1995.}
\lref\MVW{E. Martinec, {\it Phys. Lett.} 
{\bf B217} (1989) 431; {\it Criticality, 
catastrophes and compactifications}, in 
{\it Physics and mathematics of strings}, V.G. Knizhnik memorial volume,
L. Brink, D. Friedan and A.M. Polyakov eds., World Scientific 1990
\semi C. Vafa and N.P. Warner, {\it Phys. Lett.} {\bf B218} (1989) 51;
W. Lerche, C. Vafa, N.P. Warner, {\it Nucl. Phys.} {\bf B324} (1989) 427.}
\lref\AGZV{V.I. Arnold,  and S.M. Gusein-Zade, A.N. Varchenko,
{\it Singularities of differentiable maps}, Birk\"auser, Basel 1985.}
\lref\CVa{S. Cecotti and C. Vafa,
{\it Comm. Math. Phys}. {\bf 158} (1993) 569.}    
\lref\CeVa{S. Cecotti and C. Vafa, {\it Nucl. Phys.} {\bf B367} (1991) 359
\semi B. Dubrovin, {\it Comm. Math. Phys. } {\bf 152} (1993) 539.} 
\lref\Evribody{
W. Lerche and N.P. Warner, 
in {\it Strings \& Symmetries, 1991}, N. Berkovits, H. Itoyama et al. eds, 
World Scientific 1992 \semi
K. Intriligator, {\it Mod. Phys. Lett.} {\bf A6} (1991) 3543
.}
\lref\WDVV{E. Witten, Nucl. Phys. {\bf B 340} (1990) 281\semi
R. Dijkgraaf, E. Verlinde and H. Verlinde, Nucl. Phys. 
{\bf B352} (1991) 59;   
in {\it String Theory and Quantum Gravity},
proceedings of the 1990 Trieste Spring School, M. Green et al. {\it eds.},
World Sc. 1991.}
\lref\Dubun{B. Dubrovin, Nucl. Phys. {\bf B 379} (1992) 627; 
{\it Differential Geometry of the space of orbits of a reflection group},
{\tt hep-th/9303152}}
\lref\EYY{T. Eguchi, Y. Yamada and S.-K. Yang, 
{\it On the genus expansion in topological string theory},
{\tt hep-th 9405106}.}
\lref\Dubde{B. Dubrovin,
{\it Geometry of 2D Topological Field Theories}, {\tt hep-th/9407018}.}
\lref\Va{C. Vafa, {\it Mod. Phys. Lett. A} {\bf 6} (1991) 337.} 
\lref\PDo{P. Dorey, {\it Int.J.Mod.Phys.} {\bf 8A} (1993) 193.}

\lref\Wa{N. Warner, {\it $N=2$ Supersymmetric Integrable Models and
{} Topological Field Theories}, to appear in the proceedings of the 1992 
Trieste Summer School, {\tt hep-th/9301088} and further references therein. }
\lref\GKO{P. Goddard, A. Kent and D. Olive, {\it Comm. Math. Phys. }
{\bf 103} (1986) 105 .}
\lref\FI{P. Fendley and K. Intriligator, {\it Nucl. Phys.} {\bf B372} 
(1992) 533 (1992)}


\lref\BPZ{A.A. Belavin, A.M. Polyakov and A.B. Zamolodchikov,
{\it  Nucl. Phys.} {\bf B241} (1984) 333.}
\lref\MoSe{ G. Moore and N. Seiberg, {\it Comm. Math. Phys. }{\bf 123} (1989)
177;
{\it Lectures on RCFT},
in {\it Superstrings {\oldstyle 89}}, 
proceedings of the 1989 Trieste spring school, M. Green {\it et al.} eds,
World Scientific 1990  and further references therein. }
\lref\JLC{J.L. Cardy, {\it Nucl. Phys.} {\bf B270} 
[FS16] (1986) 186.} 
\lref\WN{W. Nahm, {\it Nucl. Phys.} {\bf B114} (1976) 174.}
\lref\ABF{G.E. Andrews, R.J. Baxter and P.J. Forrester, {\it J.Stat. Phys.}
 {\bf 35} (1984) 193.}
\lref\TL{P. Martin, {\it Potts models and related problems in statistical 
mechanics}, World Scientific, 1991
\semi {\it Yang Baxter Equations in Integrable Systems}, M. Jimbo edr, 
World Sc. 1989.}
\lref\Wi{E. Witten, Nucl. Phys. {\bf B 340} (1990) 281.}
\lref\DVV{R. Dijkgraaf, E. Verlinde and H. Verlinde, Nucl. Phys. 
{\bf B352} 59-86 (1991):
in {\it String Theory and Quantum Gravity},
proceedings of the 11990 Trieste Spring School, M. Green et al. {\it eds.},
World Sc. 1991.}
\lref\Wa{N. Warner, {\it $N=2$ Supersymmetric Integrable Models and
Topological Field Theories}, to appear in the proceedings of the 1992 
Trieste Summer School, {\tt hep-th/9301088} and further references therein. }
\lref\GKO{P. Goddard, A. Kent and D. Olive, {\it Comm. Math. Phys. }
{\bf 103} (1986) 105 .}
%
\lref\DFZa{Vl.S. Dotsenko and V.A. Fateev,
{\it Nucl. Phys.} {\bf B240} (1984)
[FS] 312; {\it Nucl. Phys.} {\bf B251} (1985) [FS] 691;
{\it Phys. Lett.} {\bf154B} (1985) 291 \semi
A.B. Zamolodchikov and V.A. Fateev, {\it Sov. Phys. JETP} {\bf 62}
(1985) 215; {\it Sov. J. Nucl.Phys.} {\bf 43} (1986) 657 \semi
V.B. Petkova, {\it Int. J. Mod. Phys.} {\bf A3} (1988) 2945; 
V.B. Petkova, {\it Phys. Lett.} {\bf 225B} (1989) 357; 
P. Furlan, A.Ch. Ganchev and V.B. Petkova,
{\it Int. J. Mod. Phys.} {\bf A5} (1990) 2721; 
Erratum,  {\it ibid.} 3641 \semi
A. Kato and Y. Kitazawa, {\it Nucl. Phys.} {\bf B319} (1989) 474 \semi
J. Fuchs, {\it Phys. Rev. Lett.} {\bf 62} (1989) 1705; 
J. Fuchs and A. Klemm, {\it Ann.Phys.} (N.Y.) {\bf 194} (1989) 303; 
 J. Fuchs, {\it Phys. Lett.} {\bf 222B} (1989) 411; 
J. Fuchs, A. Klemm und C. Scheich, {\it Z. Phys.C} {\bf 46} (1990) 71\semi
M. Douglas and S. Trivedi, 
{\it Nucl. Phys.} {\bf B320} (1989) 461.}
\lref\Gep{D. Gepner, {\it Nucl. Phys.} {\bf B296} (1988) 757; 
{\it Phys. Lett.} {\bf 199B } (1987) 380.} 
\lref\TN{T. Nassar, Rapport de DEA, Paris 1994.}
\lref\MVW{E. Martinec, {\it Phys. Lett.} 
{\bf B217} 431- (1989); {\it Criticality, 
catastrophes and compactifications}, in 
{\it Physics and mathematics of strings}, V.G. Knizhnik memorial volume,
L. Brink, D. Friedan and A.M. Polyakov eds., World Scientific 1990
\semi C. Vafa and N.P. Warner, {\it Phys. Lett.} {\bf B218} (1989) 51;
W. Lerche, C. Vafa, N.P. Warner, {\it Nucl. Phys.} {\bf B324} (1989) 427.}

\lref\EYY{T. Eguchi, Y. Yamada and S.-K. Yang, 
{\it On the genus expansion in topological string theory},
{\tt hep-th 9405106}.}
\lref\BALZ{D. Bernard, {\it Nucl. Phys.}  {\bf B288} (1987) 628 \semi
D. Altsch\"uler, J. Lacki and P. Zaugg, {\it Phys. Lett.} 
{\bf 205B} (1988) 281 \semi
P. Christe and F. Ravanini, {\it Int. J. Mod. Phys.} {\bf A4} (1989) 897 \semi
G. Moore and N. Seiberg, {\it Nucl. Phys.} {\bf B313} (1989) 16 \semi
M. Bauer and C. Itzykson, {\it Comm. Math. Phys.} {\bf 127} (1990) 617 \semi
Ph. Ruelle, E. Thiran and J. Weyers, 
{\it Comm. Math. Phys. }{\bf 133} (1990) 305; 
{\it Nucl. Phys.} {\bf B402} (1993) 693.}
\lref\Bax{R. Baxter, {\it J. Stat. Phys.}, {\bf 28} (1982) 1.}
\lref\JMO{M. Jimbo, T. Miwa and M. Okado, {\it Lett. Math. Phys.} 
{\bf 14} (1987) 123; 
{\it Comm. Math. Phys.} {\bf 116} (1988) 507. } 
\lref\We{H. Wenzl, 
{\it Inv. Math.} {\bf 92} (1988) 349.}
\lref\Ko{I. Kostov, {\it Nucl. Phys.} {\bf B300} [FS22] (1988) 559.}
\lref\So{N. Sochen, {\it Nucl. Phys.} {\bf B360} (1991) 613. }
\lref\And{G.E. Andrews, {\it The Theory of Partitions}, Addison-Wesley, 1976.}
\lref\DFY{P. Di Francesco and S. Yankielowicz, {\it Nucl. Phys.}
{\bf B409} (1993) 186. }


\lref\ABI{D. Altsch\"uler, M. Bauer and C. Itzykson, 
{} level-rank duality}

\vskip8truemm
\rightline{\sl{A Claude} \qquad\qquad}
\bigskip
\newsec{Introduction}
\nind 
Similar features have appeared recently in various problems 
of two-dimensional field theory and statistical mechanics. In the
simplest case based on the $su(2)$ algebra, the classification 
of ordinary conformal field theories (cft's), of $\CN=2$ superconformal
field theories or of the corresponding topological field theories, 
and the construction of integrable lattice face models have all
been found to have their solutions labelled by the simply laced $ADE $
Dynkin diagrams. This is quite remarkable in view of the fact that 
the setting of the problem and 
the techniques of analysis are in each case quite different (see
\ICMP\ or sect. 6 below for a short review and a list of references). 
Although our understanding of the same problems in   cases of higher
rank $su(N)$  is much poorer, there is some evidence that important data are
again provided by a set of graphs. 
It is the purpose of this paper to show that these graphs may be 
given a geometrical interpretation as encoding the geometry (the scalar
products) of a system of vectors --the ``roots"-- and thus enable 
one to construct the group generated by the reflections in the
hyperplanes orthogonal to these roots. 

After recalling some basic facts and introducing notations concerning
reflection groups, I shall define the class of graphs that we are 
interested in (sect.2). These graphs are a natural generalization of the 
situation encountered with the $ADE$ diagrams on the one hand, and 
with the fusion graphs of the affine algebra $\widehat{su}(N)$ on the other. 
In sect. 3, 
the adjacency matrix of such a graph is used to define a scalar 
product on a root system and thus a reflection group. Some general
properties of these groups are proved, in particular those cases
that lead to finite groups are identified; also an analogue
of the Coxeter element is defined and shown to have interesting  properties.
In sect. 4, 
some other identifications and isomorphisms of groups are derived or 
conjectured on the basis of several explicit cases. Sect. 5
discusses the cases of ``non integrally laced graphs"   and their 
connections with  subalgebras of the ``Pasquier
algebra". Finally in sect. 6, I turn to the discussion of the physical
relevance  of these graphs and groups in the various  contexts mentionned above. 
The case of 
$\CN=2$ superconformal theories and of topological field theories 
seems the most natural setting for that interpretation and 
we shall see that the graphs and groups discussed here  are actual
realizations of general results of Cecotti and Vafa, and of Dubrovin.

A short account of this work has been presented in \ICMP.

\bigskip
 Let us start with some generalities on reflection groups. 
Let $V$ denote a vector space of dimension $n$ over $\IR$ 
with a given basis $\{\Ga_a\}$. Let $\bra\ , \ \ket$ be a symmetric 
bilinear form  on $V$. We denote
\eqn\Oa{g_{ab}=\vev{\Ga_a,\Ga_b}}
and assume that  $\vev{\Ga_a,\Ga_a}=2$. This allows one 
to define the linear transformation $S_a$
\eqn\Ob{S_a\quad : \quad x\mapsto x'=x-\vev{\Ga_a,x} \Ga_a}
or in terms of the components $x_b$~: $x=\sum_b x_b \Ga_b$
\eqn\Oc{\eqalign{x'_a&=-x_a- \sum_{c\ne a} g_{ac} x_c \cr
x'_b&= x_b  \qquad {\rm if}\  	\ b\ne a\ . \cr	}}

\nind 
The following properties are easily established:
\item{(i)}
$S_a$ is involutive:  $S_a^2=\II$, and preserves the bilinear form:
$\vev{S_a x, S_a y}=\vev{x,y}$. This is in fact the reflection in 
the $(n-1)$-dimensional hyperplane orthogonal to the vector $\Ga_a$.
\item{(ii)} If the $ab$ entry $ g_{ab}=0$, 
then $S_a$ and $S_b$ commute and the product $S_a S_b$ is of order 2
$$ (S_a S_b)^2=\II \ .$$
\item{(iii)} If the entry $g_{ab}=1$, then $S_a S_b$ is of order 3; 
more generally, if $g_{ab}=
2\cos {\pi p\over q}$, with $p$ and $q$ coprime integers, 
the restriction of $g$ to the 2-plane spanned by $\Ga_a$, $\Ga_b$ 
endows it with a structure of Euclidean space: 
the two unit vectors $\Ga_a$ and $\Ga_b$ make an angle ${\pi p\over q}$, 
 hence the product $S_a S_b$ is a rotation of angle 
 ${2\pi p\over  q}$, and $S_a S_b$ is of order $q$. 

\nind Let $\GC$ be the group generated by the reflections $S_a$, 
$a=1, \cdots n$. We call {\it roots} the vectors $\Ga_a$ and {\it root system}
the set of images of the roots under the action of $\GC$. 

An important issue is to know if the group $\CG$ 
is of finite or infinite order. 
One proves \Hum\ 
that the group is finite if and only if the
bilinear form $\bra\ ,\ \ket$ is positive definite. 
At the term of the discussion, one finds 
that finite reflection groups 
are classified \Cox: beside the Weyl groups of the simple
Lie algebras, $A_p$, $B_p$, $C_p$, (the two latter groups 
being isomorphic), $D_p$, $E_6$, $E_7$, $E_8$, $F_4$ and 
$G_2$, there are the symmetry groups $H_3$  and $H_4$
of the regular icosaedron and of a regular 4-dimensional polytope, and 
the infinite series $I_2(k)$ of the symmetry groups 
of the regular $k$-gones in the plane. 
If one uses as a basis a system of simple roots, then
$\vev{\Ga_a,\Ga_b} \le 0$ if $a\ne b$ \Hum, and the bilinear
form $g_{ab}$ of \Oa\ is the so-called Coxeter matrix $\GC_{ab}$
\eqn\Od{g_{ab}=\vev{\Ga_a,\Ga_b}=-2\cos{\pi\over m_{ab}}=: \GC_{ab}  }
where the integer $m_{ab}=m_{ba}$ is the order of the product $S_aS_b$, 
$(S_aS_b)^{m_{ab}}=\II$, $m_{aa}=1$.
 \def\Gha{\hat{\Ga}}
{Note  that 
for the groups of Weyl type, 
one usually takes roots $\Gha$ normalized 
differently, with the longest having $(\Gha,\Gha)=2$ and the Cartan matrix 
with integral entries 
\eqn\Oe{{\hat C}_{ab}=2{\vev{\Gha_a,\Gha_b}\over\vev{\Gha_b,\Gha_b}}
\in \IZ\ ;}
then $\GC_{ab}$
is 
\ the {\it symmetrized} form of the Cartan matrix
\eqn\Of{ {\hat C}_{ab}=
{|\Gha_a|\over |\Gha_b|} \GC_{ab}\ ,}
with $|\Gha|:=(\vev{\Gha, \Gha})^{\oh}$. 
For future reference, also note that the adjacency matrix $\hat G$
of the Coxeter-Dynkin diagram is related to the Cartan 
matrix by
\eqn\Og{ {\hat C}_{ab}=2\Gd_{ab} -\hat G_{ab} =-2{|\Gha_a|\over |\Gha_b|} 
\cos{\pi\over m_{ab}} \ .}}

\newsec{Graphs}
\nind
The construction of graphs proceeds through generalization of two cases 
under control, namely the ordinary $ADE$ Dynkin
diagrams on the one hand and the fusion graphs
of $\widehat{su}(N)_k$ on the other. I briefly review well
known facts about these two cases.

\subsec{The $ADE$ Dynkin diagrams}
\nind The $ADE$ Dynkin diagrams are unoriented graphs that 
have the property that their adjacency matrix has a spectrum 
of eigenvalues satisfying 
\eqn\Ia{ |\gamma| <2\ .}
In fact they are, together with the orbifolds $A_{2n}/\IZ_2$, 
the only unoriented connected graphs with that property \GHJ. (The latter
orbifold graphs will be discarded in the following: from the point of view 
of lattice models, they are uninteresting as they produce no new model; 
from a different view point, more relevant here, they are not 
2-colourable, see below.) Furthermore their eigenvalues take the form
\eqn\Iaa{ \gamma^{(\lambda)} =2 \cos {\lambda\pi \over h}\ ,}
where $h$ is the Coxeter number, and the integer 
$\lambda$ takes $r$ ($=$ rank of the
$ADE$ algebra) values between (and including) $1$ and $h-1$, with
possible multiplicities. 

Note that the $ADE$ graphs are tree graphs and 
may therefore be bi--coloured. 

One may also relax the condition that the entries of the matrix under
consideration are integers and consider the class of symmetric matrices 
whose elements are of the form 
\eqn\Ib{\eqalign{
G_{aa}&=0\cr
G_{ab}&= 2\cos {\pi\over m_{ab}} \quad a\ne b\quad\cr}}
for a set of integers $m_{ab}=m_{ba}\ge 2$. (The case considered above 
was thus $m=2$ or 3.) 
One may represent such a matrix by a graph with the edge $a-b$ decorated 
by the integer $m_{ab}$. The matrix is called indecomposable if the
graph is connected. One proves \Hum\ 
that the 
only indecomposable matrices \Ib\ that satisfy \Ia\ 
 are the symmetrized forms of the adjacency  matrices $\hat G$ of the 
Coxeter-Dynkin diagrams 
of all the finite groups $A$--$I$ discussed in the introduction, 
\eqn\Ibb{ G_{ab}={|\Gha_b|\over |\Gha_a|} \hat G_{ab}\ .}
Hence they are related to the Coxeter matrices of \Od\ by
\eqn\Ibx{\GC_{ab}=2\Gd_{ab}-G_{ab}\ .}


\subsec{The $\widehat{su}(N)_k$ fusion graphs. }
\nind
Let $\GL_1,\cdots,\GL_{N-1}$ be the fundamental weights of $su(N)$.
Let $\rho=\GL_1+\cdots +\GL_{N-1}$ be the sum of these fundamental weights.
I recall that the set of integrable weights (shifted by $\rho$) of
the affine algebra $\widehat{su}(N)_k$ is 
\eqn\Ic{\CP_{++}^{(k+N)}=\{\Gl=\Gl_1\GL_1+\cdots 
+\Gl_{N-1}\GL_{N-1}|\Gl_i\ge 1,\quad \Gl_1+\cdots \Gl_{N-1}\le k+N-1\} \ .}
This set of ${k+N-1 \choose k}$ weights is a finite subset, 
the so--called Weyl alcove, 
of the $(N-1)$-dimensional 
weight lattice. We may represent it by the graph obtained 
by drawing edges between neighbouring points on this lattice, and 
orienting them along the weights of the standard $N$-dimensional representation 
of $su(N)$, i.e. along the $N$ (linearly dependent) vectors $e_i$
\eqn\Id{e_1=\Lambda_1,\qquad e_i=\Lambda_i-\Lambda_{i-1},\quad i=2,\cdots
, N-1,\qquad e_N=-\Lambda_{N-1}\ .}
This graph will be denoted by $\CA^{(k+N)}_N$ and the subscript
$N$ will be omitted whenever it causes no ambiguity. 
It  is exemplified in Fig.1 by the case of $\widehat{su}(3)$ at level 3.
Recall that the Killing 
bilinear form is such that $(e_i,e_j)=\Gd_{ij}-{1\over N}$. 
%
%
%
\fig{The graph $\CA^{(6)}
$}{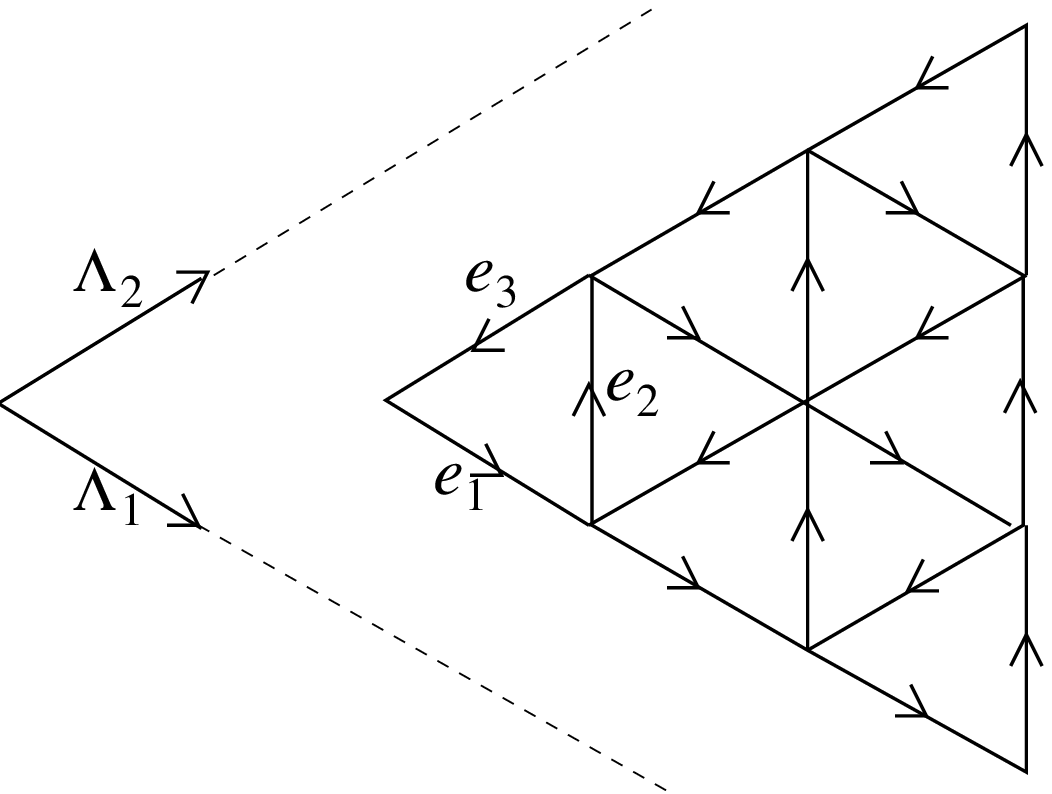}{4cm}\figlabel\triang
\def\tvp{\vrule height 2pt depth 1pt} 
\def\thp{\vrule height 0.4pt width 0.35em}
\def\cc#1{\hfill#1\hfill}

\setbox1=\vbox{\offinterlineskip
\cleartabs
\+ \thp&\cr  
\+ \tvp\cc{}&\tvp\cr 
\+ \thp&\cr  }

This graph may also be regarded as the fusion graph  of the representation
of weight $\Lambda_1$, i.e. the graph whose adjacency matrix is
$N_{\Lambda_1}\equiv N_{\copy1}$. By the Verlinde formula \Verl, we know 
how to express its eigenvalues in terms of the unitary matrix $S$
of modular transformations of the characters of the affine algebra
$\widehat{su}(N)_k$, and these expressions may be recast as
\eqn\Ie{\Gc^{(\Gl)}=\sum_{i=1}^N \exp {-2i\pi\over k+N} (e_i,\Gl)\ ,}
with $\Gl$ running over the same set $\CP_{++}^{(k+N)}$. (Since
$(e_i,\Gl) \in {1\over N}\IZ$, this reduces to \Iaa\ for $N=2$). 
{}From now on, we shall write 
\eqn\Ieaa{h=k+N\ .}

Note that the weights of $su(N)$ come naturally with a $\IZ_N$ grading, 
$\tau(.)$, the ``$N$-ality",
\eqn\Ieab{\tau(\Gl)=\sum_{j=1}^{N-1}j\Gl_j\quad \mod N\ ,}
and that the only non vanishing entries of the adjacency matrix $G_{1}$
are between  points of successive $N$-alities
\eqn\Iea{(G_1)_{ab}\ne 0 \qquad {\rm only\ if}\quad\ \tau(b)=\tau(a)+1\
\quad \mod N \ .}

\setbox3=\hbox{$\vcenter{\offinterlineskip
\+ \thp&\cr  
\+ \tvp\cc{}&\tvp\cr 
\+ \thp&\cr  
\+ $\!{}^{\vdots}$\cc{}&$\!{}^{\vdots}$\cr 
\+ \thp&\cr  
\+ \tvp\cc{}&\tvp\cr 
\+ \thp&\cr  }$}
\setbox22=\hbox{$\left.\vbox to \ht3{}\right\}$} 

\setbox2=\vbox{\offinterlineskip
\cleartabs
\+ \thp&\cr  
\+ \tvp\cc{}&\tvp\cr 
\+ \thp&\cr  
\+ \tvp\cc{}&\tvp\cr 
\+ \thp&\cr  }

\setbox4=\vbox{\offinterlineskip
\cleartabs
\+ \thp&\cr  
\+ \tvp\cc{}&\tvp\cr 
\+ \thp&\cr  
\+ \tvp\cc{}&\tvp\cr 
\+ \thp&\cr  
\+ \tvp\cc{}&\tvp\cr 
\+ \thp&\cr  }

\setbox1=\vbox{\offinterlineskip
\cleartabs
\+ \thp&\cr  
\+ \tvp\cc{}&\tvp\cr 
\+ \thp&\cr  }

\setbox33=\hbox{$\scriptstyle {N-1}$} %
\setbox34=\hbox{$\scriptstyle {p}$} %
\setbox35=\hbox{$\scriptstyle {N-p}$} %

The reason why the adjacency matrix has been labelled with a 1 
is that, when working with $su(N)$, we may, and in fact we must, generalize 
this setting and consider  the graphs associated with the fusion by 
all the fundamental representations:
$$ N_{\copy1},\ N_{\copy2},\cdots ,  N_{\copy3\copy22\copy33}\ .$$ 
The matrix considered above is thus $G_1=N_{\copy1}$, and more generally, 
for the ease of notation, we write
\eqn\If{G_p=N_{\copy3\copy22\copy34}\qquad 1\le p\le N-1 \ .}
Again by the Verlinde formula, we know that all these matrices are 
simultaneously diagonalized by the unitary matrix $S$. 
The expressions of their eigenvalues read
\eqn\Ifa{
\eqalign{\gamma^{(\Gl)}_1 
&=\sum_{i=1}^N \exp -{2i\pi\over h}(e_i,\Gl) \qquad\qquad\qquad\quad
= \chi_1(M)\cr
	 \gamma^{(\Gl)}_2  
&=\sum_{1\le i<j\le N} \exp -{2i\pi\over h}
					((e_i+e_j),\Gl)
\quad\quad =\chi_2(M) \cr
		\vdots	\quad		&\qquad\quad \vdots
\qquad\qquad \vdots \cr
 	\gamma^{(\Gl)}_{N-1} 
	&=\sum_{1\le i_1<\cdots i_{N-1}\le N}
	\exp -{2i\pi\over h}((e_{i_1}+\cdots + e_{i_{N-1}}),\Gl) \cr
   	&=\(\gamma^{(\Gl)}_1\)^* \qquad\qquad\qquad\qquad\qquad\quad\quad
	=\chi_{N-1}(M)\ .
	\cr } }
These eigenvalues may be expressed in terms of ordinary $SU(N)$ characters 
$\chi_p$ for the $p$-th fundamental representation  evaluated 
at  the $SU(N)$  matrix $M=\diag(\exp-2i\pi(e_i,\Gl)/h)$~%
\foot{An intriguing  observation is that these special matrices $M$ are
in one-to-one correspondence with the conjugacy classes of 
elements of finite order in the group $SU(N)$ \KMoPa. I am
indebted to J. Patera for this remark.}. 
The $\Gc$ are pairwise complex conjugate, $\Gc_p=(\Gc_{N-p})^*$, 
since $\sum_i e_i=0$, which reflects the fact that 
the matrices $G_p$ are pairwise transposed of one another:
\eqn\Ifb{ G_p^t=G_{N-p}\ .}
For example, in the case of $su(3)$, the graphs pertaining to the 
two fundamental representations are obtained from one another by
reversing all the orientations of edges. 

Finally, we recall the property that the set of $\Gc_p^{(\Gl)}$, 
for all $p=1, \cdots, N-1$, 
characterizes the weight $\Gl$ in $\CP_{++}^{(h)}$: 
\eqn\Ifc{{\rm if\ for\ all}\quad p=1,\cdots, N-1 \quad \Gc_p^{(\Gl)}=
\Gc_p^{(\Gm)} \quad {\rm then}\quad \Gl=\Gm\ .}
This follows from the fact that the fusion ring of  $\widehat{su}(N)$
is polynomially generated by the fundamental representations \Gepn. 
Thus under the conditions of \Ifc, the fusion matrices of $\Gl$ and
$\Gm$ are identical, which suffices to identify $\Gl$ and $\Gm$. 
%



%

\subsec{Generalized graphs}
\nind
Building upon  the two particular cases discussed above, it seems natural
to introduce a class of generalized graphs satisfying the following 
properties:

\item{1)} We are given a finite set $\CV$ of $n$ vertices, 
that are denoted $a,b,\cdots$. 
In the set $\CV$ acts an involution $a \mapsto \bar a$ 
(a generalization of the conjugation of representations). 
When working with $su(N)$, we assume that a 
$\IZ_N$ valued grading $\tau$ is assigned to these vertices, 
such that $\tau(\bar a)=-\tau(a) \ \mod N$.

\item{2)} We are also given a set of $N-1$ commuting $n\times n$ matrices, 
labelled by the fundamental representations of $su(N)$ and, 
like in \If\ above, denoted $G_p$, $1\le p \le N-1$.
These matrices have entries that are non negative integers and may thus be
regarded as the adjacency matrices of $N-1$ graphs $\CG_p$. 
Contrary to the cases encountered above, some of the edges $(ab)$ may be 
multiple, {\it i.e.} have $\(G_p\)_{ab}> 1$. 
For the sake of irreducibility, we have also to  assume
some property of connectivity of the set $\CV$. There is no 
partition $\CV=\CV'\cup \CV''$ such that 
$$\forall a\in \CV',\
\forall b \in \CV'',\ \forall p=1,\cdots, N-1,
\quad (G_p)_{ab}=0\ .$$
\nind {\it Remark}. In the same way as in \Ib, 
it is natural to extend slightly the previous 
condition and to allow non integer $G$. As we shall see in sect. 5 below, 
$(G_p)_{ab}$ of the 
specific form
\eqna\Igb
$$\eqalignno{(G_p)_{ab}&=0\  
\quad {\rm or }\quad \CP^{(N)}_p(2\cos{\pi\over h})  & \Igb a \cr
{\rm or } \qquad (G_p)_{ab}&= 
2\cos{\pi\over m_{ab}}  & \Igb b \cr
 }$$
seem to be natural choices,
with $h$ and $m_{ab}$ integers  and $\CP^{(N)}_p(x)$ a certain  polynomial of $x$ 
(eqs. (5.4-5)). By a slight abuse of language, we still
call $G$ the adjacency matrix of a graph, whose edges are decorated
by the integers 
 $m_{ab}$  or $h$. 
The considerations that follow (until sect. 5)
do not depend on the integrality of the matrix elements.

\item{3)} The edges of the graphs $\CG_p$  are compatible with the 
$\IZ_N$ grading $\tau$ in the sense that
\eqn\Ig{\(G_p\)_{ab}\ne 0 \qquad {\rm only \ if }\quad \tau(b)=\tau(a)+p\ 
\mod N\ .}
Thus for $p\ne {N\over 2}$, the edges are oriented: $\(G_p\)_{ab}
\ne 0 \Rightarrow \(G_p\)_{ba} =0$. Also, for a given pair $(a,b)$
there is at most one matrix $G_p$ with a non vanishing entry $(G_p)_{ab}$.

\item{4)} The matrices $G_p$ are pairwise transposed of one another:
\eqn\Ih{  G_p^t=G_{N-p}\ .}
Moreover each graph is invariant under the involution in the  sense that
\eqn\Ii{\(G_p\)_{\bar a \bar b}=\(G_p\)_{ba}\ .}

\item{5)}
Since the matrices $G$ 
commute among themselves, they commute with their transpose, (``normal matrices''),  
hence they are diagonalizable in a common 
orthonormal basis $\psi^{(\Gl)}$; we {\it assume} that these eigenvectors are 
labelled by integrable weights $\Gl$ of $\widehat{su}(N)$ 
and that the corresponding
 eigenvalues of $G_1$, $G_2$, \dots, $G_{N-1}$ read as in \Ifb, for
some $h$ and $\Gl \in \CP_{++}^{(h)}$; some of these $\Gl$ 
may occur with multiplicities larger than one. 

\item{6)}  We assume that $\Gr=(1,1,\cdots,1)$ is among these $\Gl$, 
with multiplicity 1: it corresponds to the eigenvector of largest 
eigenvalue, the so--called Perron--Frobenius eigenvector.

\nind
By extension of the $ADE$ case, we call ``exponents" these weights $\Gl$ 
(with their multiplicities) and denote their set by $\Exp$.
Let $\Gs$ be the automorphism of the Weyl alcove $\CP_{++}^{(h)}$
\eqn\Ij{\Gl=(\Gl_1,\cdots,\Gl_{N-1})\mapsto  \Gs(\Gl)= 
(h-\Gl_1-\cdots-\Gl_{N-1}, \Gl_1,\Gl_2,\cdots,\Gl_{N-2})\ .}
One checks that  
\eqn\Ija{\exp-{2i\pi\over h}(e_i,\Gs(\Gl))= e^{{2i\pi\over N}} 
\exp-{2i\pi\over h}(e_{i-1},\Gl) }
and thus, using \Ifa 
\eqn\Ik{\Gc_p^{(\Gs(\Gl))}= e^{2i \pi{p\over N}} \Gc_p^{(\Gl)}\ .}
There is of course another automorphism acting on $\CP_{++}^{(h)}$: 
the conjugation $\CC$ of representations
\eqn\Il{{\rm under \ }\CC\quad \Gl\mapsto \bar\Gl \qquad 
\Gc_p^{(\bar\Gl)}=\(\Gc_p^{(\Gl)}\)^*\ .}

Now consider $\Gl \in \Exp$, $\psi^{(\Gl)}$ a  
corresponding eigenvector. Define $\tilde \psi^{(\Gl)}_a=
e^{2i\pi{\tau(a)\over N}}\psi^{(\Gl)}_a$.
Then property 3) implies that 
$\tilde \psi^{(\Gl)}$ is an eigenvector of $G_p$ of eigenvalue
$ \tilde\Gc_p^{(\Gl)}=e^{2i\pi{p\over N}} \Gc_p^{(\Gl)}=\Gc_p^{(\Gs(\Gl))}$.
Thus (using \Ifc), we conclude that if $\Gl$ is an exponent, 
so is $\Gs(\Gl)$. 
On the other hand, the real matrices $G_p$ have eigenvalues that come in 
complex conjugate pairs; thus the exponents also come in complex 
conjugate pairs. 
\smallskip\nind
{\bf Proposition 1 :} {\sl The set $\Exp$ is invariant under the action of
$\Gs$ and $\CC$. Moreover one may choose}
\eqn\Ila
{\eqalign{
\psi^{(\bar\Gl)}_a 
&=\psi^{(\Gl)}_{\bar a} =
\(\psi^{(\Gl)}_a\)^*
\cr
\psi^{(\Gs(\Gl))}_a&= e^{{2i\pi\over N}\tau(a)} \psi^{(\Gl)}_a\ .\cr}}
\medskip
\nind In \DFZun\ we presented lists of solutions to these requirements
in the case $N=3$. 
\medskip
\nind {\it Remarks}\par\nind
1) For some purposes, in particular for the construction of lattice 
integrable models based on the graphs, it seems necessary to impose
the further constraint that the graph of $G_1$, say, has an ``extremal 
point", i.e. a vertex on which only one edge is ending and from which 
only one edge is starting. This constraint will not play any role 
in the following and may thus be omitted. \par\nind
2) It may be more economic to 
consider a single matrix $G$
\eqn\Ilb{G=G_1+G_2+\cdots+G_{N-1}\ .}
The edges of the graph $\CG$ it encodes connect only vertices 
of different $\Gt$. Conversely if the graph $\CG$ is given, 
and if the grading of the vertices is known, 
each matrix $G_p$ may be identified as the adjacency matrix  
of the subgraph joining pairs of vertices of $\tau$ differing by $p$.

\bigskip
For later use, I now introduce  an explicit  parametrization 
of the matrices $G_p$. 
I assume that the vertices of $\CV$ have been ordered according to
increasing $\tau$: first the vertices with $\tau=0$, then $\tau=1$, etc.
Then the matrices $G_p$ are $N\times N$  block--matrices of the  form
\eqn\Im
{\eqalign{
G_1 &=\pmatrix{ 0 & A_{12} & 0      & \cdots & 0          \cr
                0 & 0      & A_{23} &\cdots  & 0          \cr
            \vdots& 0      & \ddots  &\ddots   & \vdots     \cr
        	0 & 0      & \cdots &   0    &A_{N-1\, N} \cr
         A_{N\, 1}& 0      &        & \cdots & 0          \cr   }\ , \cr
    & \cdots \qquad \cdots\qquad \cdots\qquad \cdots \qquad \cdots 
\qquad \cdots               \cr
G_p &=\pmatrix{0  & \cdots &A_{1\,p+1}& \cdots & 0         \cr
            \vdots& 0      &          & \ddots & \vdots    \cr
		0 & 0      & \cdots   &   0    &A_{N-p\, N}\cr
     A_{N-p+1\, 1}& 0      &          &\cdots  & 0         \cr   
            \vdots& \ddots & 0        &\ddots  & \vdots    \cr
    		0 & \cdots & A_{N\, p}&        &  0        \cr   }\ ,\cdots 
 \cr}}
with the matrices $A_{ij}$ 
satisfying 
\eqn\Ima{A_{ij}^t=A_{ji}\ }
as a consequence of \Ih. 
(The matrices $A_{ij}$ are of course subject to further constraints
expressing the commutation of the matrices $G$, etc). 
Later, we shall also encounter the matrix%
\foot{Here and in the following, by a small abuse of notations,
$\un$ denotes a unit matrix, whose dimension is fixed by the context.}
\eqn\In
{T=\pmatrix{\un & -A_{12}& A_{13}& \cdots & -\Ge A_{1N}\cr
		       & \un    &-A_{23}& \cdots & \Ge A_{2N}\cr
		       & 0      & \ddots &       &    \cr
		       &        &   0   & \un    & -A_{N-1\, N}\cr
		0      &      0 & \cdots& 0      & \un \cr}\ }
where $\Ge=(-1)^N$.
It may be written as a product of upper triangular matrices in the 
two following ways:
\eqn\Io
{\eqalign{T&= \pmatrix{\un & & & & \cr
                                &\un & &0 & \cr 
				& & \ddots & &\cr
				&0 & & \un & -A_{N-1\, N}\cr
				 & & & & \un \cr}
\cdots \pmatrix{\un & -A_{12}& A_{13}& \cdots & -\Ge A_{1N}\cr
                    & \un &0 & &0 \cr
		    & & \ddots &0 & \cr
		    &0 & & \un & 0\cr
		    & & & & \un  \cr } \cr
&= \pmatrix{\un & & & & -\Ge A_{1N}\cr
             & \un & &0 & \Ge A_{2N}\cr
	     &  &\ddots & &\vdots \cr
	     &0 & & \un & -A_{N-1\, N}\cr
	     & & & & \un \cr}\cdots 
\pmatrix{\un & -A_{12}&0 & \cdots & 0\cr
            & \un & & 0 & \cr
	    & & \ddots & & \cr
	    &0 & & \un & \cr
	    & & & & \un \cr    }\ , \cr}}
which allows to write its inverse and its transpose as
\eqn\Ip
{T^{-1}= \pmatrix{\un & A_{12}& -A_{13}& \cdots & \Ge A_{1N}\cr
                    & \un &0 & &0 \cr
		    & & \ddots &0 & \cr
		    &0 & & \un & \cr
		    & & & & \un  \cr } \cdots
		    \pmatrix{\un & & & & \cr
                                &\un & &0 & \cr 
				& & \ddots & &\cr
				&0 & & \un & A_{N-1\, N}\cr
				 & & & & \un \cr} }
and
\eqn\Iq
{ T^t= \pmatrix{\un & &0 & \cdots & 0\cr
     	    -A_{21} & \un & & 0 & \cr
	    & & \ddots & & \cr
	    & 0& & \un & \cr
	    & & & & \un \cr    }
\cdots
\pmatrix{\un & & & & \cr
             & \un & &0 & \cr
	     &  &\ddots & & \cr
	     & 0& & \un & \cr
	    -\Ge A_{N1} &\Ge A_{N2} &\cdots &-A_{N\,N-1} & \un \cr} \ .}
Let us finally introduce the matrix $J$
\eqn\IIja{J=\diag(\un,\Ge\un,\un,\Ge\un,\cdots, \un)\ .}
These expressions will be useful soon. 


\newsec{New Reflection groups}
\subsec{Definition and first properties}
\nind
I now show that with these data, one may associate a reflection group
in a natural way. 

Let as above  $V$ be a $n$-dimensional space over $\IR$, with a basis
$\{\Ga_a\}$ labelled by the points $a$ of the set $\CV$. We then
introduce a bilinear form defined by the following expression, 
that depends on whether $N$ is even or odd
\eqn\IIa{g_{ab}=\vev{\Ga_a,\Ga_b} =2\Gd_{ab}+ 
\( (-1)^{N-1}G_1 +G_2 +(-1)^{N-1}G_3+\cdots +(-1)^{N-1} G_{N-1}
\)_{ab}\ , }
(or $g_{ab}=2\Gd_{ab}+ (-1)^{(N-1)(\Gt
(a)-\Gt(b))}G_{ab} $
in terms of the single matrix $G$ of \Ilb). An alternative form
is
\eqn\IIaie{g= J(T+T^t)J^{-1}}
in terms of the matrices introduced at the end of sect. 2.3. 
We then
consider the group $\GC$ generated by the reflections $S_a$.
In the case $N=2$, one recovers the expressions \Od--\Ib. We shall 
see below what are the virtues of the choice of signs in \IIa.

Note that the groups are generated by the reflections $S_a$ and that these
generators satisfy the relations $S_a^2=\II$, as well as
 $(S_aS_b)^q=\II$ under
the conditions mentionned in sect. 1. Generically, 
they satisfy also other relations (see examples below in sect. 
4.1), and thus the group
cannot be called a ``Coxeter group". We shall rather use the 
denomination ``reflection group''. 

Since we know by hypothesis all the eigenvalues of the $G_p$, 
those of the metric $g$ read
\eqn\IIaa{g^{(\Gl)}=\sum_{p=0}^N (-\Ge)^p\gamma_p^{(\Gl)}\qquad\qquad \Gl\in \Exp}
where we have extended the formulae \Ifa\ to $\Gc_0^{(\Gl)}=\Gc_N^{(\Gl)}=1$
and as before, $\Ge=(-1)^{N}$. 
It will be very useful to use a multiplicative form of this
eigenvalue
\eqn\IIab{g^{(\Gl)}=\prod_{i=1}^N (1-\Ge e^{-{2i\pi\over h}(e_i,\Gl)}) }
that follows from \Ifa. 
 
For $N=2$, as recalled above, all the graphs that satisfy the
previous constraints lead to a finite reflection group. For $N\ge 3$, 
on the contrary, the  group is generically of infinite order. 
More precisely, 
\smallskip
\nind {\bf Proposition 2}~:~ {\sl The 
form $\bra\ ,\ \ket$ 
of Eq. \IIa\ is 
definite positive if and only if }
\eqn\IIb
{\eqalign
{N=2 \qquad &\forall h\ge 3 \cr
 N=3 \qquad & h=4, 5 \cr 
 N>3 \qquad & h=N+1\ .\cr} }
\smallskip
{} To prove this, we shall  exhibit a non positive 
eigenvalue of the matrix $g$ whenever the conditions of
\IIb{} are not fulfilled. 
According to the Proposition 1,
all the images of $\rho$  under the action of $\sigma$ 
are always exponents. 
\def\Gll{\Gl_{\ell}}
For the exponent $\Gl_{\ell}:=
\Gs^{\ell}(\rho)$, 
$\ell=0,\cdots, N-1$, 
one may compute the 
eigenvalue of $g$ using \IIab\ and \Ija. One finds, 
with $\xi=\exp {2i\pi/N}$ and $q=\exp i\pi/h$
\def\ksl{(-1)^{N-1}\ksi^{\ell} }
\eqn\IIca{g^{(\Gll)}= (1+\ksl q^{1-N})(1+\ksl q^{3-N})\cdots(1+\ksl q^{N-1})
}
We now choose to look at the eigenvalue corresponding to $\ell=1$ for
$N$ odd, and to $\ell={N/2-1}$ for $N$ even
. In both cases, the resulting value of
$g^{(\Gl_{\ell})}$ reads
\eqn\IIcb{ g^{(\Gl_{\ell})} = 
-\prod_{j=0}^{N-1} 2\cos\( {1\over N} +{1-N+2j\over
2h}\)\pi \ .}
It is finally a simple matter to check that all the arguments of the cosine 
are between $-{\pi\over 2}$ and ${\pi\over 2}$, and thus $g^{(\Gl_{\ell})}
\le 0$, 
for  $N=3$, $h\ge 6$ or for $N\ge 4$, $h\ge N+2$. 

It remains to examine the cases of \IIb. 
Leaving aside the case $N=2$ which is well known, let us consider
first the cases $N\ge 3$, 
$h=N+1$. Then   the exponents take  their values among 
the integrable weights (shifted by $\Gr$) of $\widehat{su}(N)_1$, 
and again by Proposition 1, these values are all reached. 
For these exponents  
$\Gr$ and $\Gs^{\ell}(\Gr)= \Gr+\GL_{\ell}$, $\ell=1,
\cdots, N-1$, 
the  direct calculation shows that 
the possible eigenvalues of the metric $g$ are 
either $(N+1)$ or $1$, all positive. 
Finally, for the last case
of \IIb, $N=3$, $h=5$, the possible exponents are among the
six integrable weights of $\widehat{su}(3)_2$ and 
one checks the positivity of the eigenvalues of $g$ for each of them.
This establishes \IIb{}.

What are the graphs satisfying \IIb? We leave again aside the case $N=2$, 
which has already been discussed in the Introduction. In the case 
$N\ge 3$, $h=N+1$, we have seen that all the weights of level 
1 appear in the spectrum of exponents. 
The $\widehat{su}(N)_1$ fusion graphs are solutions, and it
is not difficult to prove that there is no isospectral graph
satisfying properties 1)--6) of sect. 2.3.
Finally, for $N=3$, $h=5$,  
if one takes only 
$\Gr$ and its orbit under $\Gs$ as exponents, the only graph 
with 3 vertices and that spectrum is the oriented triangle graph  $\CH^{(5)}$
of Fig. 2   with $G_{01} =G_{12}=G_{20}=2\cos{\pi\over 5}$;
if one  takes as  exponents all the 
six integrable weights of $\widehat{su}(3)_2$, i.e. the
$\Gs$-orbit of $\Gr$ and the $\Gs$-orbit of $2\Gr$, the only  graph
with a matrix of entries integral or of the form \Igb{} 
is the fusion graph $\CA^{(5)}$ of $\widehat{su}(3)_2$ 
(Fig. 2). It is very likely that one cannot take the 
second triplet of exponents  (the orbit of $2\Gr$) 
with a multiplicity different from zero or one, as will 
follow from a conjecture discussed below in sect. 4.3. If so, this completes
the list of graphs that lead to a positive definite form $g$.

For all these cases, the groups are of finite 
order, thus in the list discussed in the Introduction.
 We shall identify them below. 

\fig{Three $SU(3)$ graphs yielding a finite group.}
{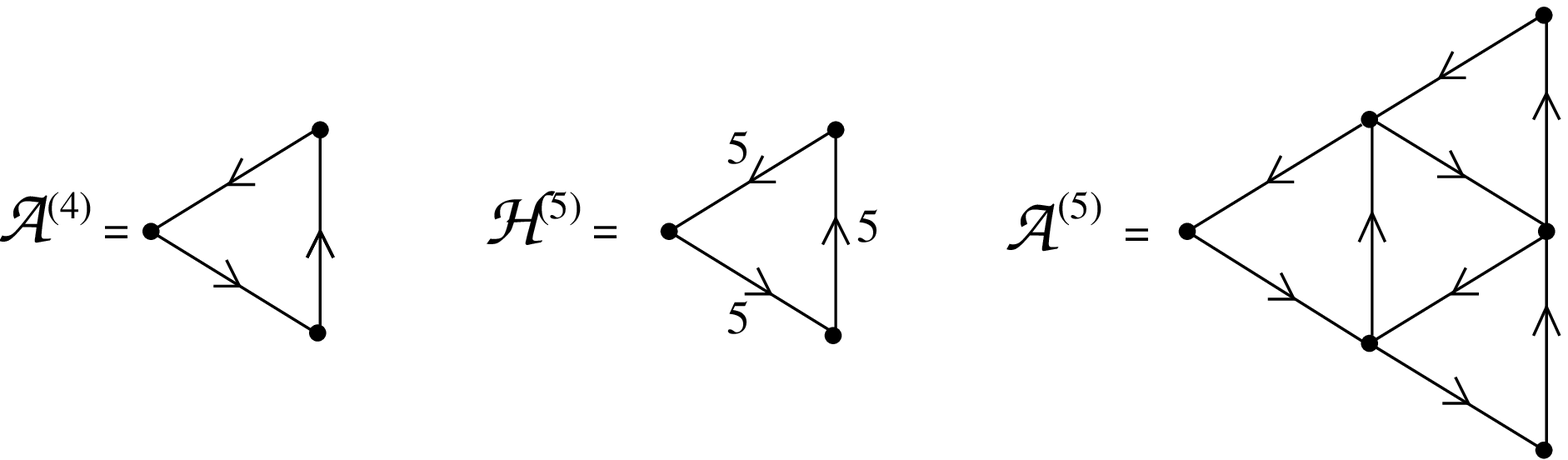}{80mm}\figlabel\sutrfin
%
%
%

%
\subsec{The Coxeter element}
\nind We now come to a non trivial property of the groups generated by this
procedure, that depends crucially on the assumptions made in sect.2 	
and on the choice of signs in \IIa. 
I first recall that in the case of finite reflection groups, 
the product of all generators pertaining to a set of simple 
roots 
$$ R= \prod_a S_a\ ,$$
called the Coxeter element, has two remarkable properties:
\item{i)} it is independent, up to conjugation, of the simple set 
and of the order of the factors;
\item{ii)} its spectrum of eigenvalues is given again by the exponents $\Gl$
in the form
\eqn\IIg{  {\rm eigenvalues \ of\ } R = \{ \exp{-2i\pi\over h} \Gl \}\ .}
\nind A weaker version of that property is still true for the groups
introduced in this paper. 

Let $R$ stand for 
\eqn\IIh{R = \prod_{\tau(a)=0} S_a  \prod_{\tau(b)=1} S_b  
\cdots\prod_{\tau(f)=N-1} S_f \ , }
i.e. the product of the blocks of reflections 
of given $N$-ality. Then
\smallskip
\nind {\bf Proposition 3 : }{\sl The element $R$ is independent
of the order  of the $S$ within each block; 
it is conjugate in the linear group $GL(n)$ to the product $-T^{-1} T^t$ of the matrices
defined in \Ip, \Iq, and its spectrum is of the form
\eqn\IIi{
 (-1)^N \exp N {-2i\pi\over h}(e_j,\Gl)\qquad \Gl\in \Exp\, , \quad j\ 
{\rm fixed: }\  1\le j \le N }
In particular this set of eigenvalues
is independent of $j=1,\cdots, N$.  } 

That the $S$ may be permuted within each block follows from the
fact that with the above assumptions, if $\tau(a)=\tau(b)$, 
then $g_{ab}=0$, hence $S_a$ and $S_b$ commute.

The proof of \IIi\ relies on a simple extension of the original proof 
by Coxeter of the analogous statement for finite reflection groups \Coxe. 
We make use of the notations introduced in \Im, \In\ to write the successive 
blocks of \IIh\ as
\eqn\IIj{\eqalign{
S_{[0]} &:=\prod_{\tau(a)=0} S_a  =
\pmatrix{-\un & \Ge A_{12} & -A_{13}& \cdots & \Ge A_{1N}\cr
	      &  \un & 0 &\cdots & 0 \cr
	      \vdots & & & & \vdots \cr
	      0 & & \cdots & & \un \cr} \cr
&= \pmatrix{\un & \Ge A_{12} & -A_{13}& \cdots & \Ge A_{1N}\cr
	      &  \un & 0 &\cdots & 0 \cr
	      \vdots & & & & \vdots \cr
	      0 & & \cdots & & \un \cr}
\pmatrix{-\un &  & & & \cr
	     & \un  & 0& & \cr
             & & \ddots & & \cr
	     & & 0 && \un \cr} =: B_0 C_0 \cr
S_{[1]}:=\prod_{\tau(b)=1}& S_b  = 
\pmatrix{\un & & & & \cr
	 & \un & \Ge A_{23}& -A_{24} & \cdots\cr
	  & & \un & & \cr
	&0 & &\ddots & \cr
	& & & & \un \cr}
\pmatrix{\un & && & \cr
	\Ge A_{21} & -\un & & & \cr
	& & \un & & \cr
	 & 0 & &\ddots& \cr
	 & & & & \un \cr} =: B_1 C_1 \cr
& \cdots \qquad   \qquad \cdots \cr
S_{[N-1]}&:=\prod_{\tau(f)=N-1} S_f  = 
\pmatrix{\un & & & & \cr
	 & \un & & & \cr
	 & & \ddots & & \cr
	 \Ge A_{N\, 1}& -A_{N\, 2} & \cdots & \Ge A_{N\, N-1} & -\un\cr}=: 
C_{N-1}\ . \cr }}

The matrices $B_p$ and $C_q$ just introduced are such that 
$B_p$ commutes with all $C_q$, for $q<p$. This
allows one to rewrite
\eqn\IIk{\eqalign{
R=S_{[0]} S_{[1]}\cdots S_{[N-1]}&= B_0C_0 B_1 C_1\cdots C_{N-1}\cr
&= B_0 B_1 \cdots B_{N-2}C_0 C_1 \cdots C_{N-1}\ .\cr}}
Then it is readily seen that 
the product $B_0 B_1 \cdots B_{N-2}$ is conjugate to the matrix $T^{-1}$
written in \Ip, and likewise, that $C_0 C_1 \cdots C_{N-1}$ is
conjugate to $-T^t$ of \Iq
\eqn\IIka{
\eqalign{J B_0 B_1 \cdots B_{N-2}J^{-1}&= T^{-1}\cr
J C_0 C_1 \cdots C_{N-1}J^{-1}&= -T^t \cr }}
where the matrix $J$ has been introduced in \IIja.
Thus we have shown that our putative ``Coxeter element" $R$ is 
conjugate to $-T^{-1}T^t$. 

\def\ve{\varepsilon}
On the other hand, let us form the polynomial $\GD(z)$ that admits the roots
$ \ve_j^{(\Gl)}:=\exp- {2i\pi\over h}(e_j,\Gl)$
\eqn\IIl{\eqalign
{\GD(z)&=\prod_{j=1}^N\prod _{\Gl\in \Exp} \(z-\ve_j^{(\Gl)}\) \cr
&=
\det\( z^N \un -z^{N-1} G_1 + z^{N-2} G_2 +\cdots +(-1)^{N-1} zG_{N-1}+ 
(-1)^N \un\)\cr
&= \det \pmatrix{
(z^N+\Ge)\un & -z^{N-1} A_{12}& z^{N-2} A_{13}& \cdots & -\Ge z A_{1N}\cr
-z\Ge A_{21}    & (z^N+\Ge)\un & -z^{N-1}A_{23}&     & \cr
\vdots & & &\ddots & \cr
-z^{N-1}A_{N\, 1} & z^{N-2}A_{N\, 2} &\cdots & &  (z^N+\Ge)\un \cr }  \cr}}
where the second expression is obtained using the formulae \Ifa;
in the third line, we have used the block decomposition \Im\ of the 
$G_p$ matrices. If we now multiply the $i$-th row of blocks by 
$z^{-i+1}$ and the $j$-th column by $z^{j-1}$, which does not
affect the determinant, 
the result depends only 
on $z^N$ and is expressed in terms of the matrix $T$ defined in \In
\eqn\IIm{\GD(z)= \det \( z^N T + (-1)^N  T^t \) = \det \(z^N \un + (-1)^N 
 T^{-1} T^t\)\ .}

Then the  ``Coxeter element'' $R$ of
\IIh\ which  is conjugate to $-T^{-1}T^t$ has  by the 
previous discussion the spectrum $\{(-\ve^{(\Gl)})^N\}$, which
completes the proof of the proposition. In this last assertion, 
we have dropped the index $j$ on $(-\ve^{(\Gl)})^N$ to emphasize 
the fact that this set of $N$-th  powers of the $\ve_j^{(\Gl)}$
does not depend on $j$. This follows from the consistency
of the argument, but 
is also a direct consequence of 
\Ija.

\medskip
{\it Remarks.}\par \penalty 10000 
Note that if we order the product in \IIh\ in the reverse way, 
as each block has square one, one gets the inverse of $R$. The 
latter, however, is also conjugate to $R$, since any $\ve^{(\Gl)}$
comes along with its complex conjugate $\ve^{(\Gl)\,*}$ in the spectrum. 
In the special case $N=3$, one can do a little more and prove the 
independence of $S$ (up to conjugacy in the group) with respect to the order 
of the three blocks. This follows once again from the fact that the
three blocks in \IIh\ have square one. 
\par
Note also that the  ``Coxeter element" $R$ is of finite order, equal 
{\it at most} to $h$ if $Nh$ is even and to  $2h$ if $Nh$ is odd. (The order
 may be smaller, e.g. for the graphs $\CD^{(6)}$ and $\CD^{(9)}$ of
Fig. 3 it is 2 resp. 6, while the value of $h$ is given by the superscript.) 
%
%
%
\fig{Two $SU(3)$ orbifold graphs.}
{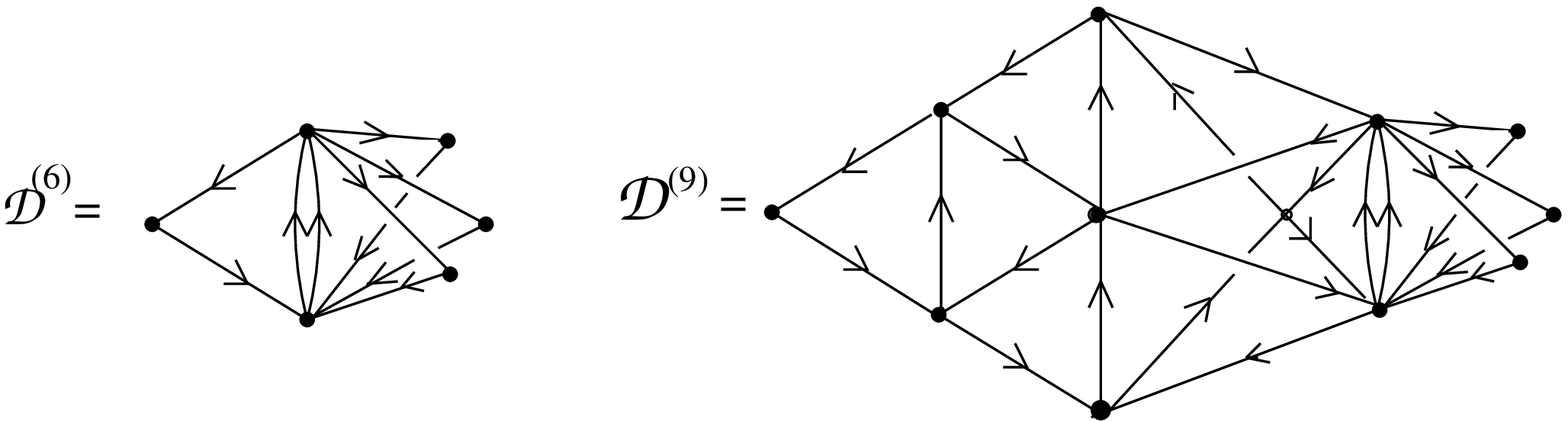}{80mm}\figlabel\suorb%
%

%
\newsec{Identification of some groups.}
\nind In this section, we shall identify some of the groups introduced
above, and establish a certain number of isomorphisms between pairs
of such groups. Given two groups $\GC$ and $\GC'$ generated by the 
reflections relative to two root bases  $\{\Ga\}$ and $\{\Ga'\}$, 
the strategy for establishing the isomorphism $\GC\cong \GC'$ is
to prove that the basis $\{\Ga'_a\}$ is found within the root system 
of the $\{\Ga_b\}$, i.e. that each $\Ga'_a$ is obtained by a finite number
of reflections of $\GC$ acting on some $\Ga_b$. This will be referred 
to as a ``change of basis within the root system". 


\subsec{Finite groups}
\nind According to the discussion of sect. 2.1, 
the group $\GC$ generated by the $S_a$
is finite for $N=3$ and $h<6$, and must therefore identify with one of the 
well known finite reflection groups. The 
groups associated with the graphs $\CA^{(4)}$,  
$\CA^{(5)}$ 
and $\CH^{(5)}$  of Fig. 2
coincide indeed  respectively with the finite groups $A_3$, $D_6$ and
$H_3$ of orders 24,  $2^5 6!$ and 120. 
This is proved by finding a different basis $\{\Gb_i\}$ of the \
space $V$ within the root system $\Ga$ 
such that the $\Gb$ are simple positive roots of the finite Coxeter group 
and that their scalar 
product is thus encoded in the conventional Dynkin diagram. 
The change of basis in the last two cases is as follows (Fig. 4)
$$\eqalign{ \CA^{(5)}\cong D_6 \quad & 
\Gb_1=\Ga_1,\ \Gb_2=S_1\Ga_2,\ \Gb_3=S_2\Ga_3,\ \Gb_4=S_2\Ga_4,\cr 
& \Gb_5=S_3S_4 \Ga_5, \ \Gb_6=S_3S_2S_5\Ga_6
\cr
\CH^{(5)}\cong H_3 \quad & 
\Gb_1=\Ga_1,\ \Gb_2=-\Ga_2,\ \Gb_3=S_2S_1\Ga_3\ .\cr
}$$ 
%
%
%
%
%
\fig{Labelling of vertices of two pairs of graphs leading to isomorphic
groups. Beware that the left one is regarded as
a $su(3)$ graph (in which the orientations have been removed)
whereas the right one is a $su(2)$ one; the 
prescription for the scalar products of roots varies according 
to Eq. \IIa.}{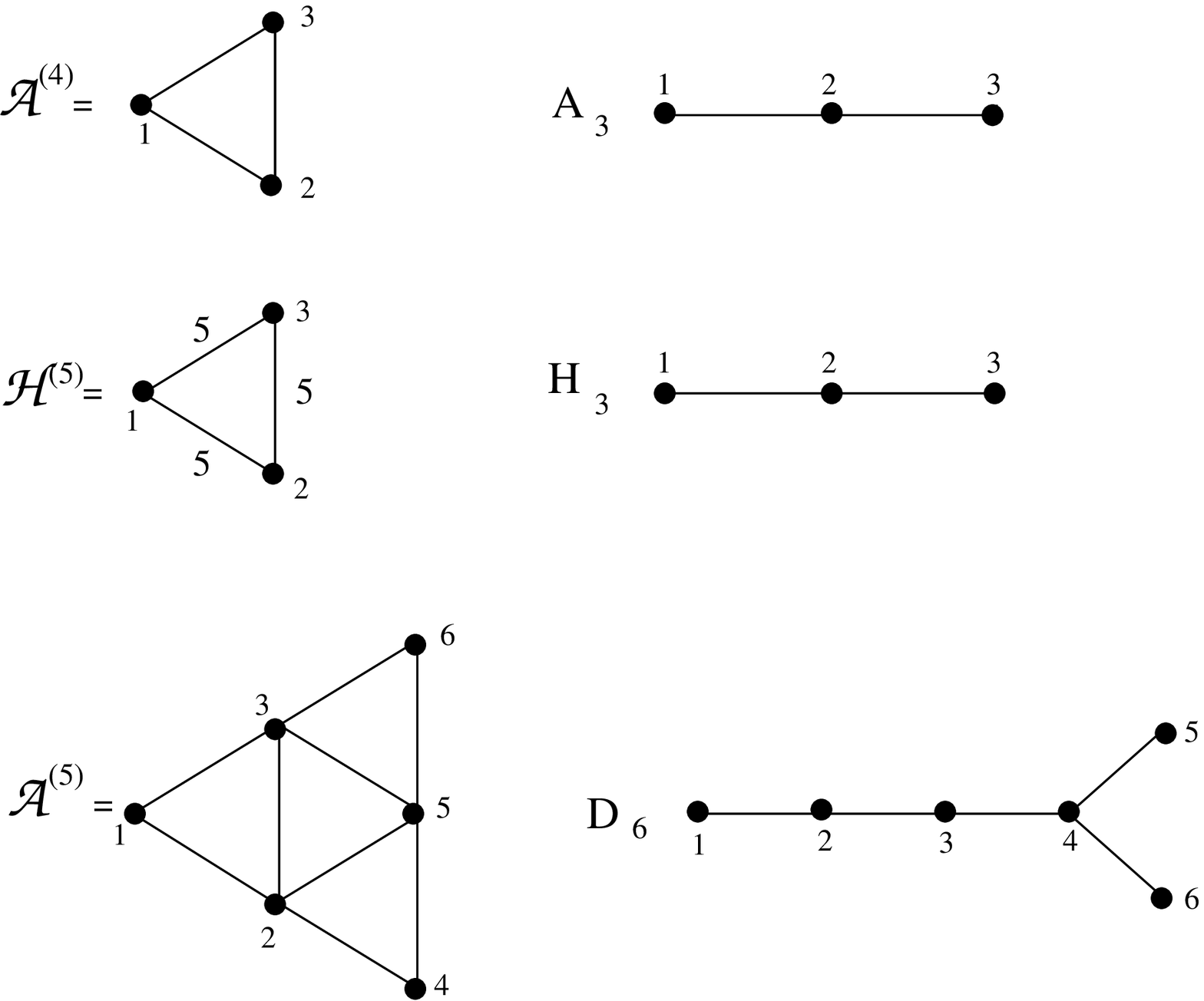}{80mm}\figlabel\chbas
The discussion of the case of $\CA^{(4)}$ may be extended 
to that of the group associated with the 
graph of weights of $\widehat{su}(N)_1$, which is nothing else than 
the finite Coxeter-Weyl group $A_N$.
For $N$ odd, $\vev{\Ga_a,\Ga_b}= (\Gd_{ab}+1)$, and
the $\Gb$ defined by
\eqn\IIs{
\Gb_1=\Ga_1,\qquad \Gb_a=S_{a-1}\Ga_a=\Ga_a-\Ga_{a-1},\ a=2,\cdots,N}
satisfy $\vev{\Gb_a,\Gb_a}=2$, $\vev{\Gb_a,\Gb_{a+1}}=-1$, and all the
other scalar products vanish. The $\Gb$ are
thus identified with the simple positive roots of $A_N$. For $N$
even, $\vev{\Ga_a,\Ga_b}=
 (-1)^{a-b} (1+\Gd_{ab})$. Then if one takes 
\eqn\IIt{\Gb_1=\Ga_1\quad \Gb_2=\Ga_2\quad 
\Gb_a=S_{a-2}\Ga_a=\Ga_a-\Ga_{a-2},\ a=3,\cdots,N}
one finds again that the $\Gb$ are the simple positive roots of
$A_N$ with a peculiar labelling 
$$ \buildrel {N-1} \over \bullet\!\!\cdots\cdots
\!\!\buildrel 3 \over \bullet \!\!{{}\over{\qquad}}
\!\!\buildrel 1 \over \bullet \!\!{{}\over{\qquad}}
\!\!\buildrel 2 \over \bullet \!\!{{}\over{\qquad}}
\!\!\buildrel 4 \over \bullet 
\!\!\cdots\cdots \buildrel N \over \bullet \ .$$

As we shall see below, the groups have a natural interpretation 
in the context of $\CN=2$ superconformal field theories. 
The first two identifications could thus have 
been anticipated from identifications between coset realizations 
of ${\cal N}=2$ superconformal theories \KS. 
Indeed (see for instance \DFLZ)
\eqn\IIr{\eqalign{
{SU(3)_1\over SU(2)_2\times U(1)} 
&\equiv {SU(2)_2\over 
U(1)}
\cr
 {SU(3)_2\over SU(2)_3\times U(1)}
&\equiv \Big[{SU(2)_8\over
 U(1)}
\Big]_{``D_6" } \ . \cr}}
The identification of the group associated with 
the graph  of weights of $\widehat{su}(N)_1$ with  
the finite reflection group $A_N$ reflects also an 
identification of ${\cal N}=2$ coset superconformal theories, namely
\eqn\IIu{{SU(2)_{N-1}\over U(1)} \equiv {SU(N)_1\over SU(N-1)_1\times U(1)}\ .}

An alternative description of these groups is by a {\it presentation} by
generators and relations \CoMo. One may check for example that the group
associated  with the graph $\CA^{(4)}$ 
is generated by the three generators $S_1,S_2,S_3$ subject to
\eqn\Iua{\eqalign{S_1^2 &=S_2^2=S_3^2=\II \cr
(S_1 S_2)^3&=(S_2S_3)^3=(S_3 S_1)^3=\II\cr
S_1S_2S_3S_2S_1&=S_3S_2S_3\ . \cr}}
The last relation (or any permutation thereof) is an example of these 
non trivial relations satisfied generically by the reflections $S_a$
of our root systems. 
Likewise, the group associated  with the graph $\CH^{(5)}$ above (Fig. 2)
is generated by the three generators $S_1,S_2,S_3$ subject to
\eqn\Iub{\eqalign{S_1^2 &=S_2^2=S_3^2=\II \cr
(S_aS_b)^2 &= S_c S_a S_c S_b S_c \cr 
(S_aS_bS_c)^2 &=S_b S_c S_a S_b \cr }}
with $a\ne b\ne c\ne a$ in the last two relations. Eq \Iub\ imply 
 $(S_aS_b)^5=\II$ as expected, and $(S_aS_bS_c)^5=\II$. 


\subsec{Generalities on the infinite cases}
\noindent
When the conditions \IIb\ 
are not fulfilled, the bilinear form $g$ is non definite positive, but one 
may still assert that the numbers of negative eigenvalues and of zeros are 
even. 
Consider an eigenvalue 
$g^{(\Gl)}$ of $g$ associated with an exponent $\Gl$. Proposition 1 
tells us that the conjugate  $\bar\Gl$ is also an exponent. 
If $\Gl\ne \bar\Gl$, as $g^{\bar\Gl}=(g^{\Gl})^*$, (see \Il), and are both real
as eigenvalues of a real symmetric form, 
 they give equal contributions
to the signature of $g$.
If $\bar\Gl=\Gl$, a close look at the expression \IIab\ of  
$g^{(\Gl)}$   
shows that its
factors come in complex conjugate pairs,  that it  
is thus non negative and that in fact it cannot vanish. 
We thus conclude that the signature of $g$ contains an even number of
zeros and an even number of minus signs.  
Note that the form is degenerate only at specific values of 
$h$ (for example $h=6,8,10,\cdots$ for $N=3$) whereas the existence of negative 
eigenvalues is the generic situation according to Proposition 2.


\subsec{Identification of some infinite cases}
\nind In this section, we shall identify  some of the groups of infinite
order associated with graphs and propose some conjectures.

\nind 
The  identity  \IIu\ is a particular case of a more
general one  that states that the $\CN=2$ theory based on
\eqn\IIuu{{SU(n+m)_{k}\over SU(n)_{k+m}\times SU(m)_{k+n}\times U(1)} }
is independent of permutations of the three integers $m,n,k$; in 
particular, taking $m=1$
\eqn\IIv{{SU(n+1)_{k}\over SU(n)_{k+1}\times  U(1)} 
\equiv {SU(k+1)_{n}\over SU(k)_{n+1}\times U(1)}\ , }
which suggests the following \par\nind
{\bf Conjecture 1:} {\sl
The reflection groups associated with
the graphs of integrable weights of $\widehat{su}(n+1)_k$ and 
$\widehat{su}(k+1)_n$ 
are isomorphic.} \par\medskip\nind
 Both graphs have 
${k+n \choose n}$ vertices, and there is a one-to-one bijection
between these vertices (i.e. weights) provided by the reflection 
of the corresponding Young tableaux along their diagonal. 
\foot{ Note that the relations between these two 
situations is {\it not} what is referred to as level-rank duality 
in the literature \Fu, 
which compares $\widehat{su}(n)_k$ and 
$\widehat{su}(k)_n$.} 

This conjecture may be verified in the case
of $\widehat{su}(4)_2 \equiv \widehat{su}(3)_3$, for which the 
following change of basis within the root system maps the graphs on one another
(the $\Gb$'s refer to $\suh(4)_2$, the $\Ga$'s to $\suh(3)_3$)
\eqn\IIva{\eqalign{
\beta_1&= \Ga_1\quad \Gb_2=\Ga_2-\Ga_1\quad \Gb_3=\Ga_4-\Ga_2+\Ga_1
\quad \Gb_4=\Ga_3-\Ga_1\quad \Gb_5=\Ga_6-\Ga_5 \cr
 \Gb_6&=\Ga_5-\Ga_4-\Ga_3+\Ga_2 \qquad \Gb_7=\Ga_8-\Ga_7+\Ga_6-2\Ga_5+\Ga_4
\qquad \Gb_8=\Ga_7\cr
 \Gb_9&=\Ga_{10}+\Ga_9-3\Ga_8+\Ga_7-3\Ga_6+4\Ga_5
+\Ga_3 -3\Ga_2+\Ga_1 \cr
 \Gb_{10}&=\Ga_9+\Ga_8-\Ga_7-3\Ga_5+\Ga_4+\Ga_3+\Ga_2-\Ga_1 \ .\cr}}
We shall see below that the conjecture is also consistent with 
further data and conjectures.
%
%
%
%
\fig{Labelling of vertices (hence of roots)  of the two graphs
of $\suh(3)_3$ and $\suh(4)_2$.   In the former, edges
code scalar products equal to $+1$. 
The latter graph is  the graph 
of matrix \Ilb, with the edges of $G_2$ that give rise 
to scalar products $=+1$ indicated by  broken lines,
those of  $G_1+G_3$ (solid lines) coding scalar products $=-1$. 
(For clarity, the Young tableau corresponding to vertex 10 has
not been depicted: it is made of three rows of two boxes.)  }
{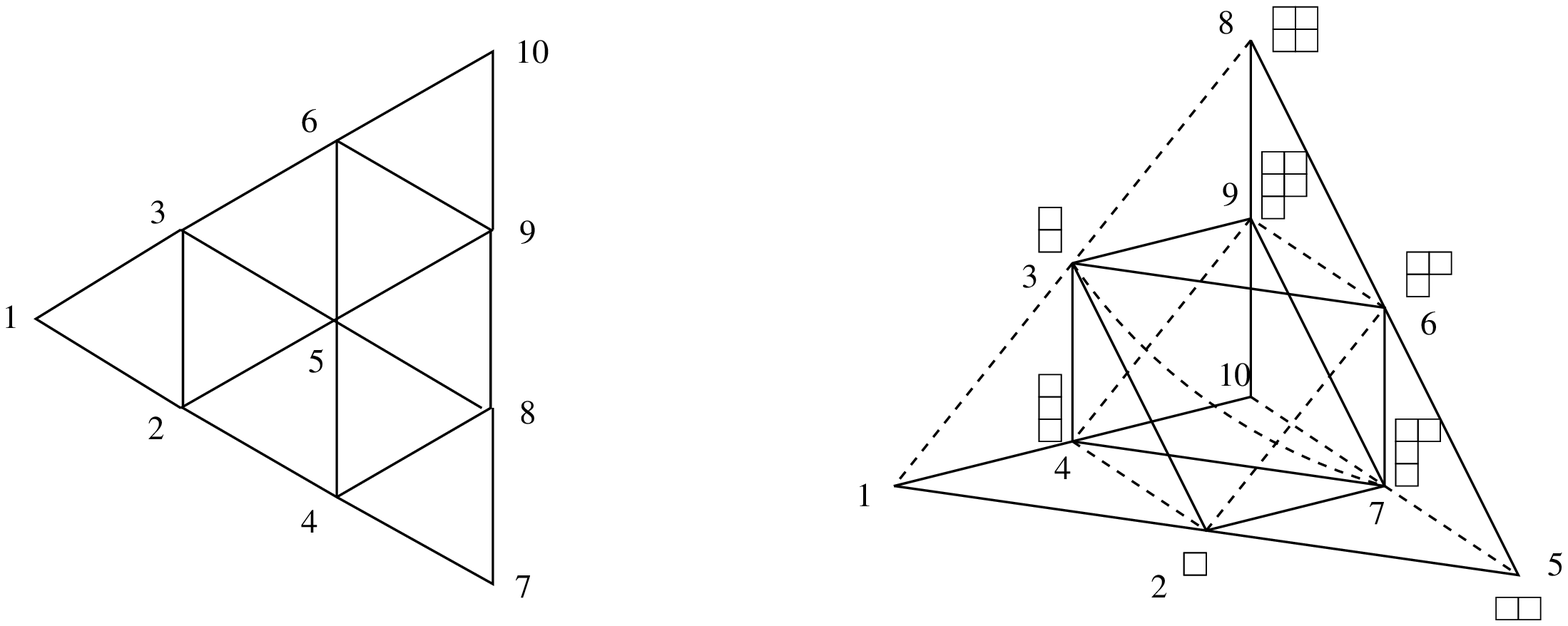}{80mm}\figlabel\chbax
Since this 
duality maps the representations of 
 $\widehat{su}(n+1)_k$ onto those 
of  $\widehat{su}(k+1)_n$ by just reversing the Young tableaux
it is clear that the number of representations with a 
given number of boxes is the same in both:
\eqn\IIw{\eqalign{{\rm card}
&\{\Gl\in \CP^{(k+n+1)}_{++}(su(n+1)), \sum_{i=1}^n
(\Gl_i-1)i=\ell\}  \cr
& ={\rm card}\{\Gm\in \CP^{(k+n+1)}_{++}(su(k+1)), \sum_{i=1}^k
(\Gm_i-1)i=\ell\}\ .\cr}
}
This leads to the  
observation that the spectra of their
``Coxeter elements" computed according to \IIi\ coincide. Indeed 
the spectrum of $R$ for
 $\widehat{su}(n+1)_k$ is
\eqn\IIx{\exp -i\pi{(n+1)(k+1)\over n+k+1} \exp {2i\pi\over n+k+1}
\sum_i\,i(\Gl_i-1) }
 and is by \IIw\ the same as that of 
 $\widehat{su}(k+1)_n$.
The same property may be checked in the other cases 
 where we have established  the isomorphism of 
two reflection groups $\GC$  and $\GC'$.  Moreover, in the cases $H_3\cong
\CH^{(5)}$, $A_N\cong \CA_{N}^{(N+1)}$, $N$ odd or $N=4$, 
one checks explicitly that the two elements $R(\GC)$  
and $R(\GC')$ are conjugate in the group. 
This  leads to the conjecture that the ``Coxeter 
element'' $R$ has (up to conjugation) a more intrinsic nature that 
suggested by the special presentation \IIh. More precisely
\par\nind{\bf Conjecture 2:} {\sl For two groups $\GC$ and $\GC'$
associated by the previous construction with two graphs $\CG$ and 
$\CG'$  of $su(N)$, resp. $su(N')$, the isomorphism $\GC\cong\GC'$
implies that the Coxeter elements $R$ and $R'$ are conjugate in the group.}

\medskip
\nind {\it Remark}. Returning to the discussion at the end of sect. 3.1, we 
thus see  that a graph with a spectrum given by the $\Gs$-orbit of $\Gr$,
and the $\Gs$-orbit of $2\Gr$ with a multiplicity larger than 1 could not 
match the spectrum of the Coxeter element of any of the finite
Coxeter groups.

\medskip
In the rest of this section, we shall see that some groups of infinite
order associated with graphs may be identified with groups
encountered in singularity theory. 

This is the case of the group associated with the graph 
corresponding to the weight lattice of $\suh(3)$ at level 3. 
At level 3, (\ie $h=6$), the graph depicted on Fig. 1 has 10 vertices. 
By a suitable change of basis of the $\Ga$'s within the root system, 
%
%
it is seen that 
the group is isomorphic 
to the monodromy group of the singularity $X^6+Y^3+a X^2Y^2$
tabulated as $J_{10}$ in \AGZV. 
This may not be obvious on the appearance of the generalized Dynkin diagram 
describing the intersection form of the vanishing cycles of the 
$J_{10}$ singularity (Fig 6). 
%
%
%
%
\fig{The Dynkin diagram of  the $J_{10}$ singularity. Here 
the double broken line one is coding a scalar product equal to $+2$, 
the solid ones a scalar product equal to $-1$. }
{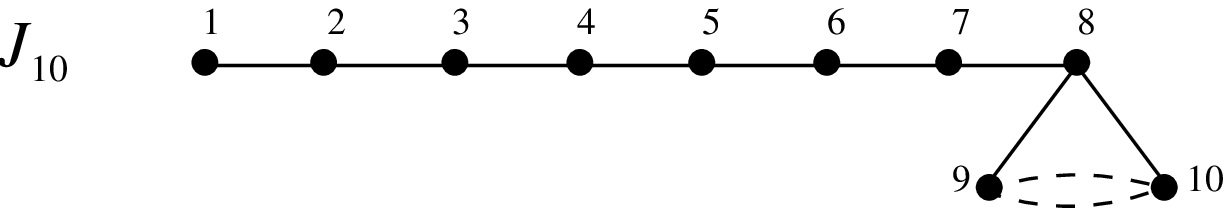}{80mm}\figlabel\jdix
Note that this singularity is also often associated with the 
Dynkin diagram of the affine algebra $\hat E_8$.
(The apparent mismatch between the rank 9 of $\hat E_8$ 
and the number 10  of vertices of $J_{10}$ 
reflects the extension of the Cartan algebra of the former by an
additional independent generator dual to the central element.)

That this graph has to do with this  singularity
is no surprise, as the  polynomial $X^6+Y^3+a X^2Y^2$
is (for a specific value of the coefficient
$a$) the homogeneous part of the fusion polynomial of $\suh(3)$
at level 3 \Gepn. This leads to the natural 
\par \nind {\bf Conjecture 3:} {\sl 
The group associated with the graph of weights of $\suh(N)_k$ is 
the monodromy 
group of the singularity described by the homogeneous part of the 
fusion polynomial, i.e. the terms of
degree $k+N$ in $t$ in the expansion of 
$\ln (1-t X_1+t^2 X_2+\cdots +(-t)^{N-1}X_{N-1})$.}
\par\nind  This conjecture  may be 
established for $N=3$, because for polynomials in only two variables, 
one can make use of the method of A'Campo \AC\ 
which simplifies greatly the determination
of the monodromy group whenever one has a 
resolution of the singularity 
 such that: i) all critical points are real, 
ii) all the saddle point values vanish. 
 As shown by Warner  \Nick, one may find such a resolution 
of the homogeneous 
part of the fusion potential of $\suh(3)$ at all levels.
Then the method of A'Campo provides us
with a description of the monodromy group by a generalized 
Dynkin diagram which upon a change of basis within the root system 
may be recast in the form of the weight lattice of 
$\suh(3)_k$ \Nick   (Fig~7). 
%
%
%
\fig{A generalized  Dynkin diagram equivalent to the graphs of figures
\chbax\ and \jdix. Same convention as in Fig. \chbax.  }
{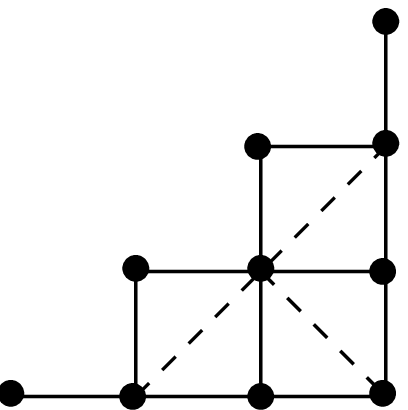}{20mm}\figlabel\acamp

This conjecture may be shown to be consistent with the 
previous one on the rank-level duality. Consider the pair
$$ \suh(N)_{N-2} \ ,
\, \suh(N-1)_{N-1}\ . $$
Although the fusion potential of $\suh(N)_{N-2}$ has one more variable 
than the one  of $\suh(N-1)$, the extra variable $X_{N-1}$ 
appears at most quadratically in it and does not affect the singularity
theory. The two monodromy groups must thus be isomorphic, 
which agrees with Conjecture 1.

\newsec{Non integrally laced graphs}
\nind This section is devoted to a closer study of the situation
where we allow some of the matrices $G_p$ to have non integral entries. 
This is a generalization of what was encountered 
with the classical finite groups not of $ADE$ type, see the end
of sect. 2.1. It must be stressed
that all the considerations of this section are based on empiric 
observations, a good understanding of which is still missing.

We shall first recall an observation made in \Zub\  where it was
shown that the non $ADE$ Coxeter-Dynkin diagrams appear in the discussion of
algebras associated  with the $ADE$ ones. 
Let us return to the normal matrices $(G_p)_{ab}$ satisfying the properties 
listed in sect. 2.3 and assumed to have integral entries, 
and let us introduce their orthonormal eigenvectors 
$\psi^{(\Gl)}_a$. We use these eigenvectors to construct the two following 
sets of numbers
\eqna\IIIa
$$\eqalignno{
M_{\Gl\Gm}^{\ \ \Gn}&=\,\sum_{a\in \CV}\, {\psi_a^{(\Gl)}\psi_a^{(\Gm)}
\psi_a^{(\Gn)\,*}
\over\psi_a^{(1)}} &\IIIa a \cr
N_{ab}^{\ \ c}&=\sum_{\Gl\in \Exp} {\psi_a^{(\Gl)}\psi_b^{(\Gl)}
\psi_c^{(\Gl)\,*}
\over\psi_1^{(\Gl)}} &\IIIa b \cr
} $$
where we have assumed the existence of a selected vertex denoted 1 
such that none of the components $\psi_1^{(\Gl)}$ vanishes. These
two sets of numbers may be regarded as structure constants of two
commutative associative algebras attached to the graph $\CG$. For the
graphs of type $\CA$ (the truncated weight lattices), these two algebras
 are 
isomorphic and reduce to the fusion algebra. I call the first algebra 
$M$ the Pasquier algebra \VP. 
Now, in all cases, we may regard the numbers $N_{ab}^{\ \ c}$
as the entries of matrices $N_{ab}^{\ \ c}=(N_a)_b^{\ c}$, 
and write
\eqn\IIIab{\eqalign{ \forall \Gl\in \Exp,\quad \forall p&=1,\cdots, N-1\qquad 
\sum_a (G_p)_{1a} \psi^{(\Gl)}_a  = \Gc_p^{(\Gl)} \psi_1^{(\Gl)} \cr
{\rm i.e.}\qquad \Gc_p^{(\Gl)} &=
\sum_a (G_p)_{1a}{\psi_a^{(\Gl)} \over \psi_1^{(\Gl)}} \cr 
{\rm and\ hence}\qquad 
G_p &= \sum_a (G_p)_{1a} N_a\ . \cr\ }}
Thus the matrices $G_p$  are 
linear  combinations of the $N_a$'s    with non negative 
integral coefficients. 
In fact in many cases, the vertex ``1" is connected (in the sense
of $G_1$)  to a single vertex $a$ and hence $G_1=N_a$. In those  cases, the 
matrices $N$ of \IIIa{b} provide  an actual realization of an idea 
of Ocneanu and Pasquier \Ocun\VP\ to look at 
the associative algebra attached to the vertices of the graph 
and generated by $G_1$. 

Now, start from a graph $\CG$, 
compute its $M$ algebra  and 
look for possible subalgebras of the latter consistent with the
requirements of sect. 2.4. In other words, we look for a subset 
$\widetilde{\Exp}$ of the set  $\Exp$ of exponents such that
\eqn\IIIb{\Gl,\Gm\in \widetilde{\Exp},\quad M_{\Gl\Gm}^{\ \ \Gn}\ne 0 
\Rightarrow \Gn\in \widetilde{ \Exp}}
and we demand that the set $\widetilde{\Exp}$ satisfies the same 
properties as $\Exp$, namely that it is stable under the 
action of $\Gs$ and $\CC$ and contains the weight $\Gr$ of
the identity representation.  Call $\widetilde{ M}_{\Gl\Gm}^{\ \ \Gn}$ 
the new structure constants obtained by restricting 
${ M}_{\Gl\Gm}^{\ \ \Gn}$ to 
$\Gl,\Gm,\Gn\in \widetilde{\Exp}$. 
It is easily seen that they are diagonalized by a set of 
$\tilde\psi_a$ 
$$ \tilde\psi_a^{(\Gl)}={\psi_a^{(\Gl)}\over \CN_a}$$
for a subset of $a\in \CV$ and a suitable
normalization $\CN_a$.  Use these $\tilde\psi$  to construct the dual algebra 
of matrices  $\tilde N_{\Ga}$.  In general, the entries of the 
$\tilde N_{\Ga}$ are irrationals.  
In \Zub\ it was noticed that when this procedure is applied to
the $ADE$ Dynkin diagrams, among these matrices $\tilde N$, 
one at least, call it $\tilde G$, has non negative entries of the form
$\tilde G_{\Ga\Gb}=2\cos{\pi\over m_{\Ga\Gb}}$, 
$m_{\Ga\Gb} \in\{2,3,\cdots,\} $,
and in fact is the symmetrized form 
\Ibb\ of the adjacency 
matrices of a non $ADE$ Coxeter-Dynkin diagram! 
In that way, the Dynkin diagram $B_n$ is obtained from 
$A_{2n-1}$, $C_n$ from $D_{n+1}$, $F_4$ from $E_6$, $G_2$ from $D_4$, 
$H_3$ from $D_6$ and $H_4$ from $E_8$. The only 
exception is the case of $I_2(k)$ where the $\tilde N$ algebra
is generated by the two-by-two matrices $\tilde N_1=\II=
\pmatrix{1&0\cr 0 &1\cr }$ and $\tilde N_2=
\pmatrix{0&1\cr1&0\cr}$ whereas the Coxeter matrix 
\Ib\ is $(2\cos{\pi\over k}) \tilde N_2$. 
\foot{As shown by Shcherbak, Moody and Patera \SMP, there is a simple
way to see how the Coxeter groups of type $B\cong C,F,G,H$ or $I$ 
appear as subgroups of some $ADE$ Coxeter group and
how an appropriate folding of the $ADE$ Dynkin 
diagram gives rise to the Coxeter-Dynkin diagram of the subgroup.
The two previous constructions are not independent, as
will be explained elsewhere.}

Remarkably, the same procedure applies to the known solutions 
with integral $(G_p)_{ab}$ to the conditions of sect. 2.4,   
that have a set of non negative structure constants of
the $M$ and $N$ algebras. It 
produces more solutions with non integral entries and 
manufactures subgroups out of the original groups. We may call
these new solutions ``non integrally laced'' graphs resp.  groups, 
to refer to the generic non integrality of the entries of the matrices. 
As in the case $N=2$, two classes of solutions emerge. In the first
class,  among the
matrices $\tilde N$ of the dual subalgebra, one can find at least 
$N-1$ matrices that have non negative entries of the form
$2\cos{\pi\over m_{\Ga\Gb}}$, 
$m_{\Ga\Gb} \in\{2,3,\cdots,\infty\} $. Moreover these matrices 
(or linear combinations with integral coefficients thereof) 
 qualify as possible matrices $G_p$, i.e. satisfy the conditions 
of sect. 2.3 (cp. \IIIab). This is 
exemplified on three cases on Fig. 8 using three graphs
found in \DFZun. These graphs were shown to be associated with
modular invariants coming from 
conformal embeddings of $\suh(3)_k$ in a larger algebra (resp.
$\suh(3)_5 \subset \suh(6)_1$,
$\suh(3)_9 \subset (\hat e_6)_1$,
and $\suh(3)_{21} \subset (\hat e_7)_1  $) \ChRa; the subalgebra of the Pasquier algebra
is labelled by the weights belonging to  the block of the identity. 
%
%
%
\fig{Three graphs of $su(3)$ (on the left) having a $M$-subalgebra
that leads to new graphs (on the right). The dotted line denotes an
edge carrying $\sqrt{3}=2\cos{\pi\over 6}$, and the double
lines are indeed edges with $G_{ab}=2$.}
{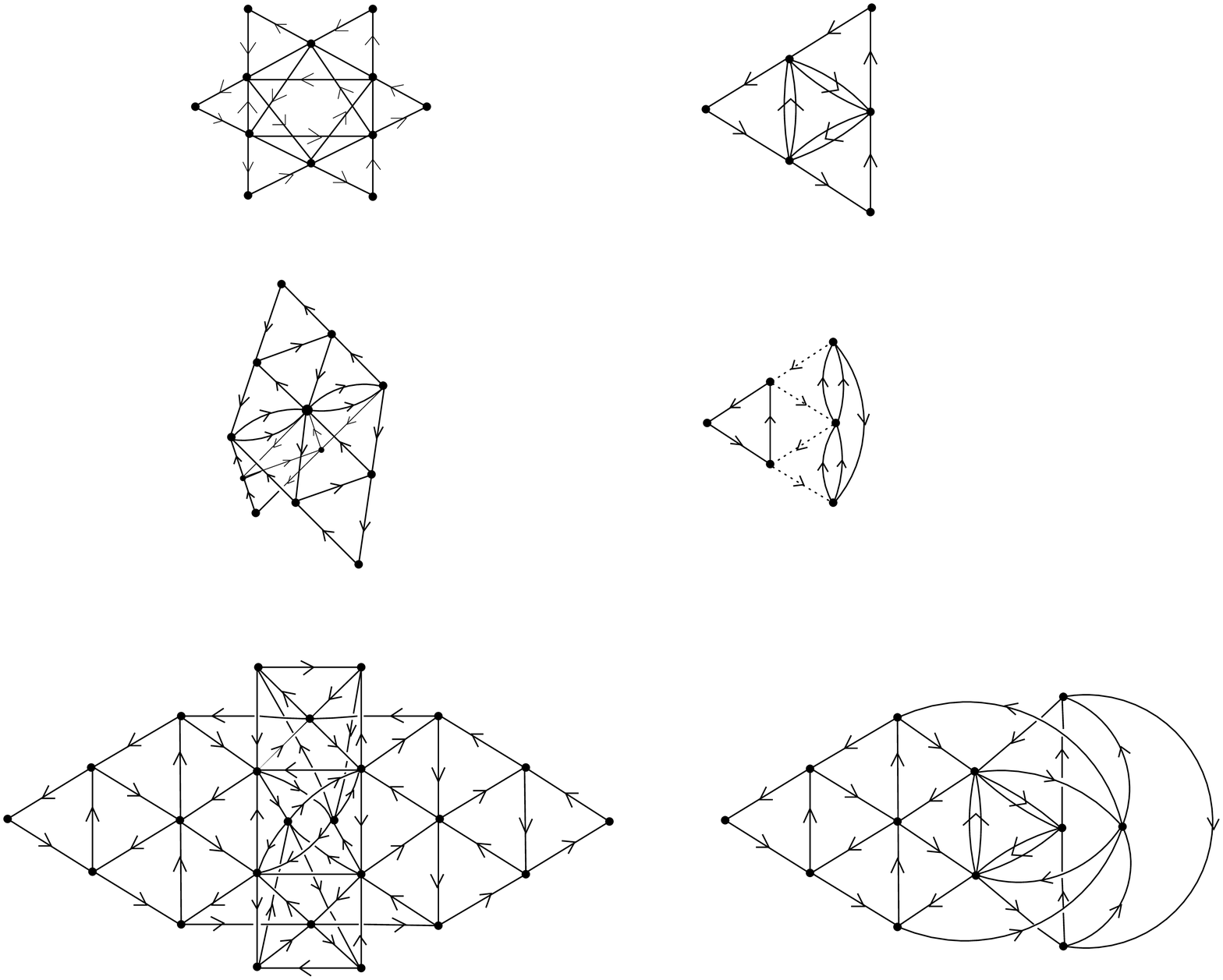}{100mm}\figlabel\subgr
The second class of solutions is a simple generalization of the situation
discussed above with the Coxeter graph $I_2(k)$. For any graph relative 
to $N\ge 2$, the Pasquier $M$ algebra admits a subalgebra whose 
generators are labelled by the $\Gs$-orbit of $\Gr$: $M_{\Gr}$, $M_{\Gs\Gr}$,
\dots,$M_{\Gs^{N-1}\Gr}$. This is in fact the smallest subalgebra consistent
with our requirements. 	In that case, the procedure of considering 
the dual subalgebra does not produce the right adjacency matrices 
but rather the adjacency matrices $G_p$ of $\suh(N)_1$. Those have a spectrum 
of eigenvalues equal to $N$-th roots of unity, $\xi^{\ell p}$, 
$\xi:= \exp {2i\pi\over N}$, $\ell=0,\cdots,N-1$, instead of the 
eigenvalues $\Gc_p^{(\Gr)} \xi^{\ell p}$, which follow from 
\Ifa. The correct adjacency matrices are thus of the form
\eqn\IIIe{(G_p)_{ab}= \Gd_{a+p, b\ \mod N}\, \Gc_p^{(\Gr)}, \qquad a,b=0,
\cdots N-1\ ,}
where $\Gc_p^{(\Gr)}$ may be expressed as a $q$-deformed binomial coefficient,
$q=e^{i\pi/h}$
\eqn\IIIf{\Gc_p^{(\Gr)}= {N\choose p}_q
={\sin {\pi N\over h}\cdots \sin {\pi (N-p+1)\over h}
\over \sin {\pi \over h}\sin {2 \pi \over h}\cdots \sin {p\pi \over h}} \ .}
One may prove that the latter expression is a polynomial 
of degree $p(N-p)$   in $2\cos {\pi\over h}$. 
An example of this class is
provided by the graph $\CH^{(5)}$ exhibited in Fig. 2   for which 
the  non-vanishing entries of the adjacency matrix $G_1$ are
$(G_1)_{01}=(G_1)_{12}=(G_1)_{20}=1+2\cos{2\pi\over 5}=4\cos^2{\pi\over 5}-1
(= 2\cos {\pi\over 5})$. 

Clearly a systematic analysis and classification of the possible
solutions would be highly desirable. We shall see below how these
Pasquier algebra and  subalgebras appear in the context of conformal field theories.

\newsec{Physical interpretation of the graphs and reflection groups}

\subsec{In conformal field theories or lattice models}
\nind It has been mentioned in the Introduction that there are empirical 
relations between our graphs and the classification
of $su(N)$ integrable lattice models and 
of $su(N)$ WZW or coset conformal field theories. 
Recall that in the context of lattice models 
one looks for solutions of the Yang-Baxter equation based on the 
quantum group $\CU_q(sl(N))$. One
finds classes of solutions indexed by  graphs of the type discussed above 
\VPun\DFZun\DFZ. The
 r\^ole of the graph $\CG_1$ is to 
specify what are the allowed configurations of the degrees 
of freedom (or ``heights"). The Boltzmann weights of the model 
are obtained by finding a representation of a quotient of the
Hecke algebra (a deformation of the algebra of the symmetric group)
on the space of paths on the graph $\CG_1$. 
In the context of cft's, the r\^ole of the graph is more indirect. 
The partition functions of these 
theories on a torus are sesquilinear forms
in characters $\chi_{\Gl}$  
 of the affine algebra $\widehat{su}(N)_k$ 
indexed by integrable 
weights $\Gl$. The constraint of modular 
invariance and of non-negativity of the coefficients 
restricts enormously the possible expressions $\sum_{\Gl,\Gl'}N_{\Gl\Gl'}
\chi_{\Gl}\chi_{\Gl'}^*$. 
It turns out that in many instances, diagonal terms in these 
expressions are labelled by $\Gl$ running over one of the sets 
of ``exponents'' encountered above in the spectrum of integrally laced graphs. 
This correspondence is known to be one-to-one for $\suh(2)$ theories,
for which all the modular invariants may be labelled by an $ADE$
Dynkin diagram \CIZ. For $\suh(N)$ theories, $N>2$,  however, some modular 
invariants exhibit diagonal terms that do not satisfy the property
of invariance of the set $\Exp$ under $\Gs$ that followed from 
condition 3) imposed in sect. 2.3 on the graphs \refs{\Itz{--}\Gan}.
It seems, however, that these cases may always be recovered 
from the others by a twist of the right sector with respect 
to the left one by an automorphism of the fusion algebra. 
Another observation is that there is a class of modular invariants
relative to the coset theories 
$$ {SU{(N)}_{k-1}\times SU{(N)}_1\over SU{(N)}_{k}\ }\ , $$
the ``minimal $W_N$ models", for which the problem does not seem 
to occur: see \PZnew\ for a further discussion.

Conversely, it has been known for some time that some graphs 
satisfying the conditions of sect. 2.3 
are irrelevant for the classification of modular invariants and
do not seem to support an integrable lattice model. Indeed
in \DFZ, some graphs had to be discarded. It thus appears that   we
are still missing some further restriction on the graphs. 

These little discrepancies notwithstanding, it seems that there is a
hidden connection between the  problems of classification
of graphs, of cft's and of integrable lattice models. 
At this point, it may be useful to recall that the manifestation 
of this connection  goes beyond 
the mere coincidence of spectra of ``exponents" of 
graphs with the diagonal terms of modular invariants and involves 
the Pasquier algebras introduced in sect. 5. 
First there is an empirical correspondance between the graphs 
that have a pair of non negative $M$ and $N$ algebras and the cft's 
whose partition function is a sum of squares of combinations of
characters. Moreover there is some 
evidence that the pattern of algebras and subalgebras of type
$M$ is connected with the structure of the operator product 
algebra (OPA) of conformal field theories and lattice models. It is 
in fact in that context that these algebras were first introduced \VP, and
the quantitative role of the $M$ algebra in the determination of
the  structure constants 
of $su(2)$ conformal theories  has been clarified recently \PZ. 
In a more recent work \PZnew, the extension of these considerations
to higher rank cases is discussed: it is shown that 
the vanishing  of the matrix elements of the $M$ algebra implies 
that of the OPA structure constants. Therefore a subalgebra of
the Pasquier algebra signals a subalgebra of the OPA. 

A few more facts are known about the association between a cft and a graph.
Starting from a graph with non negative $M$ and $N$ algebras, 
an empirical reconstruction of the modular invariant using 
the theory of ``c-algebras" \BI\  has been developed in \DFZ. 
It is also believed that the non simply (or ``non integrally") laced graphs 
do not lead to an acceptable modular invariant partition function, but 
rather to an invariant under a subgroup of the modular group \Zub. 
Reciprocally, in a variety of cases of cft's, 
(typically orbifolds or conformal embeddings), 
considerations on the OPA that generalize those of \PZ\ enable one 
to determine an appropriate graph \PZnew.

In yet another approach, these graphs have been used in a recent work
to construct invariants of three-manifolds, \`a la Turaev-Viro \Ocn. 

In view of these connections of graphs with cft's and  lattice models, it is 
natural to wonder whether 
the reflection groups constructed in this work manifest themselves in 
those physical contexts. I have unfortunately 
no definite answer to that question. The previous discussion suggests
that the reflection group might be hidden in some features of the OPA.

As already mentioned in \ICMP, this suggests a programme based on the
assumption that the previous observations are of general validity. If 
one could ascertain the connection between the 
consistent subalgebras of the OPA and reflection groups, one
might consider classifying the relevant  reflection groups, find
their presentation by generators subject to the conditions of sect. 2, and
reconstruct the graph; one would then discard the cases of 
non integrally laced graphs, that correspond to theories that are
inconsistent at higher genus. This would yield a set of admissible
graphs, and with the methods  of c-algebras or of counting 
of ``essential paths'' \Ocn, 
one would reconstruct the modular invariant partition functions. 
It remains to see how realistic this programme is \dots

\subsec{In $\CN=2$ superconformal theories and topological field theories} 
\nind The context of $\CN=2$ superconformal 
theories and of topological field theories (tft's) seems the  most 
relevant one for the interpretation of the reflection groups. 
We first recall that there is a large class of $\CN=2$ theories
amenable to a description by an effective Landau-Ginsburg (LG)
superpotential \MVW. The latter is  a quasihomogeneous 
polynomial $W$ in some chiral superfields $X_i$, $i=1,\cdots, n$, 
with an isolated critical
point at the origin in field space. In the simplest cases, it 
is thus to be found in the lists of singularities \AGZV. 
This is in particular the case
of the so-called minimal $\CN=2$ theories, based on
$sl(2)$, that are all described in that way and for which the
relevant singularity is a simple one (with no modulus), i.e. 
falls once again in an ADE classification \AGZV. 

In all cases, the elements of the chiral
ring $\CR$ are in one-to-one correspondence with those of 
the local ring of the singularity, i.e. the polynomial ring 
$\IC[X_1,\cdots,X_n]$ 
quotiented by the ideal generated by 
the derivatives $\partial W/\partial X_i$. The $U(1)$ charges $q_j$
of the chiral fields are proportional to the homogeneity degrees of 
the elements of a basis of the local ring, 
with the proportionality factor fixed by the requirement that 
$q(W)=1$. If $c$ denotes the central charge (of the Virasoro algebra),
the  $U(1)$ charges of the Ramond ground states are
$q_j^{{\rm (Ramond)}}= q_j^{{\rm (chiral)}}-{c\over 6}$.
On the other hand, it is 
a standard practice in singularity theory to look at deformations
of the polynomial that resolve the singularity and to study the
intersection form of the vanishing cycles and the monodromy group
of these cycles when the deformation parameters are changed along loops. 
For $n$, the number of variables, even, 
the intersection form is encoded in a generalized Dynkin diagram. 
Cecotti and Vafa \CVa\ have shown that this intersection form counts the 
(signed) number of solitons $A_{ab}$ interpolating between the vacua of the 
$\CN=2$ {\it supersymmetric} theory obtained by perturbing the original 
$\CN=2$ {\it superconformal} theory. For a special choice of deformation 
and of labelling, the matrix $A$ may be taken upper triangular. 
The monodromy operator is the form 
$H=SS^{-t}$ (with the notations of \CVa), with $S=\II -A$, and its
eigenvalues  are $\exp 2\pi i q_j^{{\rm Ramond}}$.  

This applies in particular to the $\CN=2$ theories based on the
cosets 
\eqn\VIza{{SU(N)_{k}\over SU(N-1)_{k+1}\times  U(1)} }
 (cp \IIv). 
There the LG potential $W$ is the quasihomogeneous part of
the fusion potential of $\suh(N)_k$; as already discussed 
in sect. 4, it is a polynomial of degree $h=k+N$ in $X_1,\cdots, X_{N-1}$ 
of respective degrees $1,\cdots, N-1$. The chiral fields may 
be labelled by integrable weights $\Gl\in \CP^{(k+N)}_{++}$ (cp \Ic). 
 Their $U(1)$ charges  are thus
\eqn\VIa{q_ {\Gl}={1\over h} \sum_{i=1}^{N-1} i(\Gl_i-1)\qquad {\rm with}
\quad \sum_i\Gl_i\le h-1\quad \Gl_i\ge 1\ .}
with $h=k+N$. The central charge $c$ of the $\CN=2$ theories under discussion
 is 
\eqn\VIb{ c  
=3 (N-1)(1-{N\over h}) \ }
thus
\eqn\VIc{q_{\Gl}^{{\rm (Ramond)}}=q_{\Gl}-{c\over 6}={1\over h} \sum_{i=1}^{N-1} 
i(\Gl_i-1)- \oh (N-1)+{N(N-1)\over 2h} \ . }
One thus sees that the eigenvalues of the monodromy operator $H$
coincide 
with those of the opposite $-R \approx T^{-1} T^t $ of the 
``Coxeter'' operator     of the present paper, (Prop. 3),  
given by \IIx:
\eqn\VId{ {\rm Eigenvalues\ of\ } (-R) =(-1)^{N-1} 
\exp\{ i\pi {N(N-1)\over h} +{2i \pi\over h } \sum_{i=1}^{N-1} 
i(\Gl_i -1)\}= e^{2i\pi q_{\Gl}^{{\rm (Ramond)}}}\ .}
In fact, 
the upper triangular $S$ of \CVa\ identifies with the conjugate by $J$ 
(eq. \IIj) of our $T$ of \In, $S=JTJ^{-1}$, 
so that our operator $-R$ identifies (up to conjugation by $J$)
with the transpose $H^t$ of their monodromy operator.

 The previous discussion has been implicitly dealing with the 
graph $\CA^{(k+N)}_N$ and the corresponding group. 
{} To make the connection with the preceeding section, we have 
been considering a $\CN=2$ superconformal theory (and its deformation)
whose genus-one partition function is constructed out of the 
diagonal modular invariant of $\suh(N)_k$.  One may also
consider other, non diagonal, modular invariants, and the resulting 
$\CN=2$ coset theory. In many cases, however, for $N>2$, the theory does not
possess a LG superpotential. The simplest example is provided by
the $SU(3)_3/\(SU(2)_4\times U(1)\)$ coset theory in which one chooses the 
orbifold modular invariant for the numerator. Correspondingly, we
take the graph $\CD^{(6)}$ of Fig. 3. Then the counting of chiral fields as 
exposed by the partition function, or the $U(1)$ charges 
\VIa\ computed from the exponents of that graph,
are incompatible with the Poincar\'e 
polynomial that would follow from a LG superpotential.

Cecotti and Vafa, however, have been able to extend their discussion to 
cases where the LG picture does not apply
. The same results hold true: the operator $H$ is the 
monodromy matrix around the origin of the solution to a linear 
system, whose consistency equations are the $tt^*$ equations \CeVa, 
and $S$ is its Stokes matrix.  Also they considered the matrix 
$B=S+S^t$ and prove (under some assumptions)
that the number $r$ (resp. $s$) of positive (resp. negative) 
eigenvalues of $B$ is
\eqn\VIe{ r=\#\{2p-\oh < q^{{\rm Ramond}} < 2p+\oh\}
\qquad s=\#\{2p+\oh < q^{{\rm Ramond}} < 2p+{3\over 2}\}\ .}
In view of the previous identification, their matrix $B$ is
nothing else than our metric $g$ (cp eq. \IIaie). 
\eqn\VIf{   B= S+S^t=J(T+T^t)J^{-1}=g}
Thus eq. \VIe\ is a statement on the signature of our metric, which
may be verified on the expression on \VIc\  for graphs
with low values of $N$ and $k$.

{} To summarize, we have found that the graphs discussed in this 
paper yield actual realizations of those discussed by Cecotti and Vafa, 
in the specific case of the $\CN=2$ theories \VIza. It is 
most likely that they  
describe the pattern of solitons that arise when the theory is
perturbed by the least relevant operator: this is the 
\Che\ perturbation, so called because in the simplest case of
$SU(2)$ theories of $A_n$ type, it changes the homogeneous superpotential
into the \Che\ polynomial $T_{n+1}(X)$ \FI. It is indeed known 
that for that perturbation, 
the pattern of solitons reproduces the classical Dynkin diagrams
or their generalizations \Evribody\DFLZ.

\medskip
The situation is quite parallel in the case of the topological 
field theories. Those may be obtained by the ``twisting"  of
$\CN=2$ theories but may also be defined and studied for their
own sake. Then the defining equations in genus zero are the
so-called Witten-Dijkgraaf-Verlinde-Verlinde (WDVV) equations \WDVV. 
Recently, Dubrovin has been able to reformulate these
equations in a coordinate independent way, as expressing the existence 
of a geometrical structure on the moduli space of these theories. 
First he proved that he could associate such a structure, i.e. a
solution to the WDVV equations with the space of orbits of
any finite Coxeter group \Dubun. 
This accounts for all the ``simple" topological
theories, i.e. with a central charge of the corresponding $\CN=2$
theory $c<3$, or alternatively, such that all $U(1)$ charges of 
moduli   be positive.  In that way, he recovered not only the $ADE$ solutions 
once again, but also others associated with the non simply
laced Coxeter matrices. 
(The consistency of the latter theories
at higher genus when coupled to gravity has been questioned recently \EYY.) 
Dubrovin also showed the existence of two independent flat metrics on the
moduli space of tft's, and, in a subsequent work, he 
 studied the differential equations that express
the flat coordinates for the first metric in terms of those of the
second one \Dubde. He proved that this differential system has a non trivial
monodromy, which under certain assumptions, is
generated by reflections. 

In the special cases of tft's described by a LG potential \Va,
this monodromy group is the monodromy group of the singularity.
We have seen that this is also the case with the groups  studied 
here (Conjecture 3). 
It is thus a very natural conjecture that the groups studied 
in the present paper are actual realizations of the  considerations
of Dubrovin for those tft's that emanate from $su(N)$ $\CN=2$ theories.



\newsec{Discussion and Conclusion}

\nind In this paper I have shown that graphs that have appeared recently
in various contexts of mathematical physics have the natural
interpretation of encoding the geometry of a root system and allow
one to construct  reflection groups.

{}From the mathematical point of view,  
in addition to all the conjectures that have been proposed, 
this paper has left many questions unanswered.
{}To quote a few:  
\item{$\bullet$} 
We have found a certain number of isomorphisms of reflection groups. 
Under which conditions on two graphs does one get two isomorphic groups?
\item{$\bullet$} 
In the finite cases like the graphs $\CA^{(4)}$ or $\CA^{(5)}$
of $su(3)$, what is the specific property of the choice of roots 
$\Ga$ within the root system of the $A_3$ or $D_6$ Coxeter groups? 
\item{$\bullet$} 
Given a reflection group, when can we assign it to a $su(N)$
graph? If the Coxeter element has an intrinsic meaning, 
(as suggested in Conjecture 2), and may be  identified in the group, 
it gives some information about  the spectrum $\Exp$ of the graph, but it remains
to reconstruct the graphs with a given spectrum. 
\item{$\bullet$} 
Can one classify these groups/graphs?
\item{$\bullet$} 
What can one say about the nature of the possible
invariants of these groups? 
Does the set $\Exp$ encode any information about 
those, as it does in the $su(2)$ cases?

\medskip Another problem  deserves a special mention.  
For each of the graphs considered in this paper  
there exists a family of matrices  $V^{\Gl}_{ab}$ 
intertwining its adjacency matrices $G_p$ 
with those --denoted here $A_p$-- of the basic fusion graph of sect. 2.2
with the same value of $h$, 
i.e. satisfying
\eqn\VIIa{ (A_p)_{\Gl \Gm} V^{\Gm}_{ab}= V^{\Gl}_{ac}(G_p)_{cb}}
(with summation over repeated indices).
In fact an explicit formula may be given for a class of $V$:
\eqn\VIIb{V^{\Gl}_{ab}=\sum_{\Gm\in \Exp} {\phi^{(\Gm)}_{\Gl}\over
\phi_{\Gr}^{(\Gm)} }  \psi^{(\Gm)}_a \psi^{(\Gm)\, *}_b\ .    }
Eq. \VIIa\  may be recast as recursive equations for the 
$V$'s which show that these coefficients are integers. The surprise, 
however, is that they are {\it non negative} integers. This was 
checked case by case in \DFZun, and then Dorey, in the $su(2)$  cases,
was able to derive it from properties of the root system of the
$ADE$ algebras \PDo. It would be interesting to see if these considerations
extend to $su(N)\,,\ N>2$ using the root systems of the present paper.
(For a physical interpretation of these coefficients in terms of 
boundary conditions see \DFZun\PDo.)

\medskip

{}From the physical point of view, there are still several missing links. 
\item{1)} In spite of many hints, we have no general  proof that the 
graphs that satisfy  the 
constraints of sect. 2.3, possibly supplemented by some additional 
conditions, encode the data relative to  modular invariants. 
\item{2)} In spite of many hints, we have no general  proof either 
that the graphs that satisfy  the 
constraints of sect. 2.3, possibly supplemented by some additional 
conditions, support a representation of the appropriate Hecke algebra and thus 
yield an integrable lattice model.  
\item{3)} In both contexts, the r\^ole of the reflection group has remained 
elusive, although there is a suggestive matching of these groups and 
their subgroups with the algebras and subalgebras of Pasquier type, that are
known to be related to the structure of the Operator Product Algebra. 
\item{4)} In the contexts of $\CN=2$ and topological field theories, 
there are many indications but no general proof that the graphs are 
a particular case of those introduced by Vafa and Cecotti, and that the
groups are those considered by these authors and Dubrovin in the study 
of monodromy problems.

\medskip
Another interesting question is to extend what has been done
here for $su(N)$ algebras to other simple algebras.
On the one hand, it is known that appropriate graphs are again
closely related to modular invariants and should permit the construction
of lattice models. See for example some graphs relevant for $G_2$ in \DFZ. 
On the other hand, not all simple algebras give rise to $\CN=2$ 
superconformal field theories \KS. It is thus likely that the
interpretation in terms of reflections groups is less developed
in those latter cases. 

\medskip


{\bf Acknowledgements} 

 This paper is dedicated to the memory of Claude Itzykson, 
who was my mentor and my friend, and without whom I wouldn't be
what I am. Claude was the touchstone of my papers, -- of the 
few that I didn't write with him~--, demanding ever more
clarity, more precision. He saw only the 
beginning of the present work and made as usual several judicious 
 comments and there is no doubt
that his criticisms and suggestions would have helped me to 
improve this paper greatly, both on  the matter and on the presentation. 
\medskip
The possibility to interpret the graphs of \DFZ\ as root diagrams
was considered several
years ago in conversations with P. Dorey, to whom I express my thanks for
his constant interest and encouragement. 
Several gaps in the present work have been 
filled thanks to the energic and friendly help of M. Bauer and N. Warner.
 I  have benefited a lot from conversations with 
B. Dubrovin. It is also a pleasure to thank D. Bernard,
Ph. Di Francesco, R. Dijkgraaf, T. Eguchi,  T. Gannon, P. Mathieu, 
H.  Ooguri, V. Pasquier, J. Patera and  V. Petkova for comments and 
suggestions.

\listrefs
\end